\documentclass[preprint,12pt]{elsarticle}
  
\pdfoutput=1
\usepackage{epsfig} 
\usepackage{graphicx}
\usepackage[ansinew]{inputenc}
\usepackage{epstopdf}
\usepackage{psfrag}
\usepackage{amsmath}
\usepackage{amssymb}
\usepackage{float}
\usepackage{multirow}
\usepackage{lineno}
\usepackage{setspace}
\usepackage{flushend}
\usepackage{array}
\usepackage{styles/tabu}
\usepackage{bm}
\usepackage{color}
\usepackage{url}
\usepackage{natbib}
\usepackage{fixltx2e}
\newcolumntype{L}[1]{>{\raggedright\let\newline\\\arraybackslash\hspace{0pt}}m{#1}}
\newcolumntype{C}[1]{>{\centering\let\newline\\\arraybackslash\hspace{0pt}}m{#1}}
\newcolumntype{R}[1]{>{\raggedleft\let\newline\\\arraybackslash\hspace{0pt}}m{#1}}
\usepackage[figuresright]{rotating}
\usepackage{subfigure}
\usepackage[title,titletoc,toc]{appendix}
\usepackage{mathrsfs}
\usepackage{amsfonts}
\usepackage{svgcolor}
\usepackage{styles/ieeetrantools}
\usepackage{makecell}
\usepackage{booktabs}
\usepackage{fixmath}
\usepackage{scalerel}

\journal{Applied Thermal Engineering}

\begin{document}
	
\bstctlcite{bibliography:BSTcontrol}

\begin{frontmatter}

\title{Modelling and cooling power control of a TES-backed-up vapour-compression refrigeration system \footnote{© 2020. This manuscript version is made available under the CC-BY-NC-ND 4.0 license \url{https://creativecommons.org/licenses/by-nc-nd/4.0/}. The link to the formal publication is \url{https://doi.org/10.1016/j.applthermaleng.2019.114415}}}

\author[us]{David Rodríguez}   
\author[us]{Guillermo Bejarano\corref{mycorrespondingauthor}}
\cortext[mycorrespondingauthor]{Corresponding author}
\ead{gbejarano@us.es}
\author[us]{Manuel Vargas}
\author[INESC_ID]{João M. Lemos}
\author[us]{Manuel G. Ortega}

\address[us]{Departamento de Ingeniería de Sistemas y Automática, Escuela Técnica Superior de Ingeniería, Universidad de Sevilla (España)}
\address[INESC_ID]{INESC-ID, Instituto Superior Técnico, Universidade de Lisboa (Portugal)}

\begin{abstract}

{\emph{
		This work addresses the modelling, power control, and optimization of a thermal energy storage (TES) system combined with a vapour-compression refrigeration facility based on phase change materials (PCM). Given a novel design of a PCM-based TES tank and its interconnection with an existing refrigeration system, the joint dynamic modelling is first studied, exploring the different time scales that coexist at the interconnected system. Diverse operating modes are defined, according to the intended use of the TES tank as a cold-energy buffer to decouple cooling demand and production, whereas the static characteristic and power limits are calculated and show the high coupling between the main cooling powers involved (TES charging/discharging power, and direct power production at the evaporator). In this light, a decoupling control strategy is proposed, where the low-level controllers are simply PI regulators and the refrigerant/secondary mass flows are considered as virtual manipulated variables, applying a feedforward-based cascade strategy. The control performance is evaluated through a thorough simulation that includes all operating modes, where the reference tracking is shown to be fast and reliable enough to address high-level scheduling strategies, where the references on the main cooling powers are intended to be imposed considering economic and efficiency criteria. }}

\end{abstract}


\end{frontmatter}


\begin{itemize}
	\item Declarations of interest: none
\end{itemize}

\section{Introduction} \label{secIntroduction}

Refrigeration and freezing using vapour-compression cycles represent an increasing item in global and economical balances in both developed and developing countries. Mainly industrial processes, but also commercial and domestic applications, involve a great deal of energy devoted to cooling \cite{Rasmussen2005}. Some reports point out that about 30\% of the total consumed energy around the world is related to Heating, Ventilating, and Air Conditioning (HVAC) \cite{jahangeer2011numerical}. Regarding domestic consumption, around 28\% of the overall consumed energy in USA is linked to refrigerators and HVAC \cite{ruz2017hybrid}.

This huge consumption amply justifies the need for studying and improving energy efficiency of current vapour-compression refrigeration systems, mainly from an economic point of view, but environmental concerns are also involved. Increasing use of environmental-friendly refrigerants, along with novel and more efficient heat exchanger design, variable-speed compressors, and electronic expansion valves (EEV) have contributed in the last decades to efficiency enhancement. Moreover, the application of Automatic Control to continuously drive the manipulated variables in order to achieve the control objectives despite changing operating conditions improves adaptability and reliability. In addition to the satisfaction of the cooling demand, that represents the main control objective, achieving as high as possible energy efficiency may be addressed through diverse techniques, which recent works have compared and analysed from the point of view of controllability and achieved energy efficiency in steady state \cite{Bejarano2017,bejarano2017suboptimal}.

However, a novel line of research is focusing in recent years on cold-energy management, in addition to optimal cooling power production. A promising strategy consists in integrating thermal energy storage (TES) systems within the refrigeration system, in such a way that a thermal reservoir may store excess cold energy and release it when necessary, similarly to strategies applied to solar thermal plants \cite{rubio2018optimal,navas2018optimal}. As a consequence, cooling production and demand may be decoupled, which implies some advantages regarding energy efficiency and operating economic cost. On the one hand, operating conditions may be loosened, lower capacity systems may be used and, more importantly, they may be used more efficiently, thereby reducing energy consumption \cite{dincer2002bthermal}. If a thermal reservoir stores a certain cold-energy surplus generated in low-demand periods, it might not be necessary to oversize the system to deal with peak-demand periods \cite{maccracken2004thermal}, which also helps reducing investment cost in the design stage. On the other hand, from an economic point of view, the addition of storage elements allows the scheduling of the cooling production according to the energy market. It is then possible to benefit from low-priced periods to produce more cold energy than required, store the surplus and release it during high-priced time slots, avoiding or at least reducing cold-energy production in such periods (\emph{peak-shifting}) \cite{dincer2002bthermal,rismanchi2012energy}. Nevertheless, this economic strategy must be compatible with the satisfaction of the cooling demand in real time. 

Regarding the design of the cold-energy reservoir, phase change materials (PCM) stand out from sensible-heat ones as the most appealing choice, due to their convenient thermodynamic properties. Besides their higher heat capacity, that allows the storage of larger amount of energy in a given volume, their temperature does not vary significantly while remaining in latent zone \cite{mehling2008heat}. This fact boosts heat transfer, since the temperature difference between the cold source and the fluid to which they transfer heat remains uniform throughout the heat exchanger. Some reviews provide classifications of both commercial and developing PCM suitable for cold-energy applications, according to melting temperature and heat capacity \cite{oro2012review}. Regarding the layout of the storage tank, the PCM encapsulation/geometry, and the design of the heat exchanger where cold energy is transferred from and to the PCM, some complete reviews detail a wide variety of solutions, where packed bed technology is one of the most successful configurations \cite{verma2008review,dutil2011review}. 

\begin{figure}[htbp]
	\centerline{\includegraphics[width=7cm,trim = 40 160 33 130,clip]
		{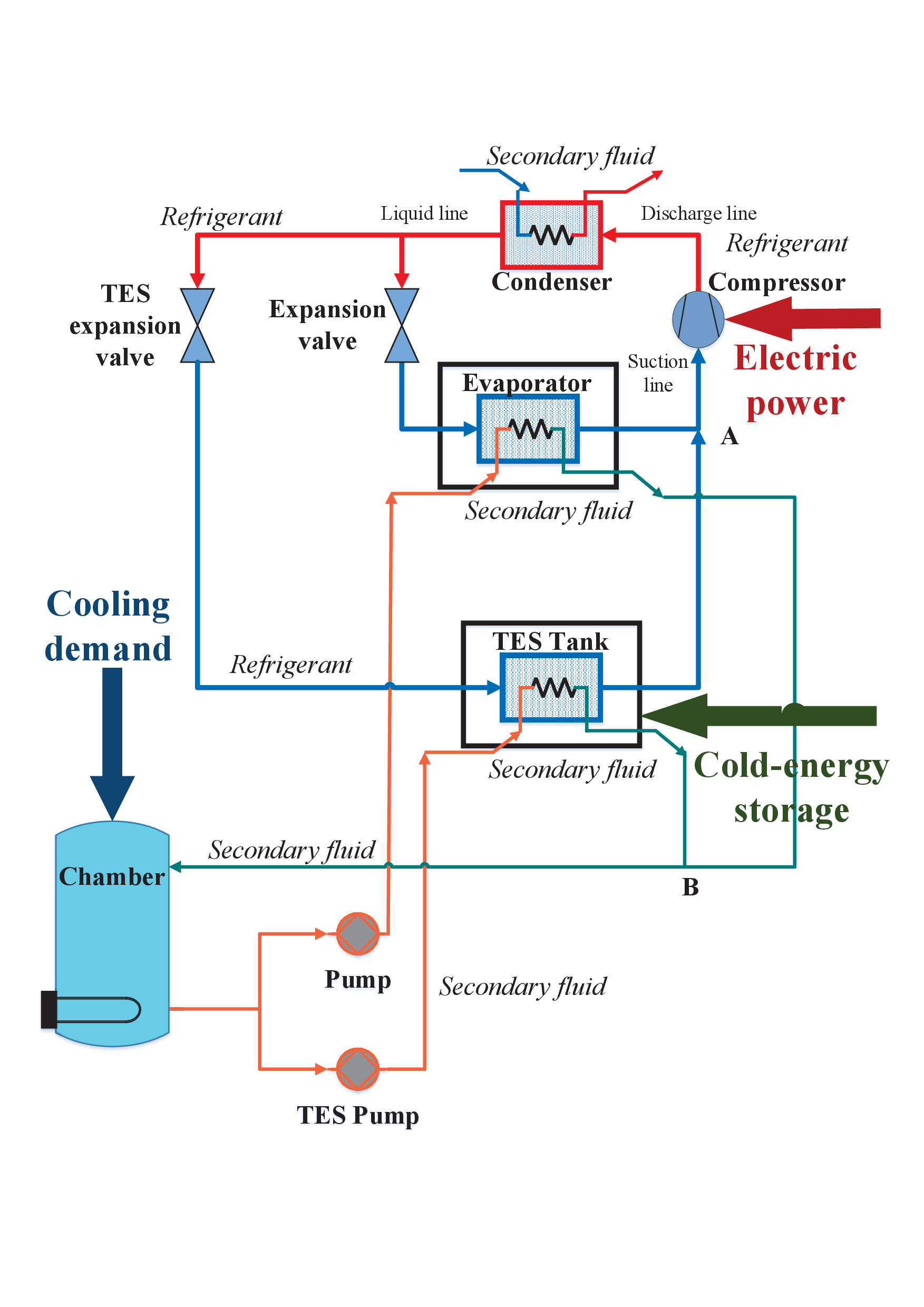}}
	\caption{Schematic diagram of the configuration proposed including a TES tank in parallel with the evaporator of a vapour-compression cycle.}
	\label{figEsquemaCicloPCM}
\end{figure}

In this article a novel hybrid configuration including a PCM-based TES tank that complements an existing vapour-compression refrigeration facility is modelled and analysed. This configuration is schematically described in Figure \ref{figEsquemaCicloPCM}. In this application, the chamber to be refrigerated consists of a tank filled with a so-called secondary fluid. The latter is pumped from the chamber to the evaporator and to the TES tank, where it is cooled and recirculated to the tank. The thermal load at the chamber is simulated by using an electric resistance. Regarding the TES tank operation, the vapour-compression cycle is intended to charge the TES tank while satisfying the cooling demand (by removing heat from the secondary fluid at the evaporator), then the cold heat transfer fluid (HTF) corresponds to the refrigerant during the charging stage. Nevertheless, when discharging the TES tank, the aforementioned secondary fluid is also intended to be cooled when circulating through the TES tank, thus the warm HTF does not correspond to the refrigerant, but to this secondary fluid. That is a key difference with respect to the typical packed bed technology, where the same HTF at different temperature is used to charge and discharge the storage tank. In this novel application, the TES tank is filled with PCM cylinders, bathed in a certain liquid, so-called intermediate fluid, while two different bundles of pipes, corresponding to the refrigerant and the secondary fluid, run through the tank, being also dipped in the intermediate fluid.

Concerning the setup, since the TES tank is arranged in parallel with the evaporator, combined cooling power can be provided to the secondary fluid, one part directly at the evaporator and another one by discharging the TES tank; this allows the TES-backed-up refrigeration system to provide higher cooling power than that achievable by using only the vapour-compression cycle, and thus this setup allows the system to face peak-demand periods maybe infeasible for the refrigeration cycle by itself. 

The main contributions of the work are highlighted next. Firstly, the modelling of the TES-backed-up refrigeration system is addressed, considering previously developed dynamic models of both the refrigeration cycle and the TES tank separately \cite{Bejarano2017,Bejarano2017NovelSchemePCM,bejarano2018efficient}. Then, the interconnection of both submodels is studied, that represents a challenge since the coupling between both subsystems is very strong in some operating modes, as detailed later. This is one of the main contributions of the work, given that the joint modelling is first addressed here. The different time scales arising in the combined system are analysed and a detailed model is presented, that describes the faster dynamics caused by the refrigeration cycle, as well as the slower dynamics arising from the heat transfer within the TES tank between the intermediate fluid and the PCM cylinders.

There are up to eight operating modes resulting from all possible combinations of activation state of the three main cooling powers involved: that provided to the secondary fluid at the evaporator, the TES tank charging power, and the TES tank discharging power. A scheduling strategy that computes the references on these three cooling powers is intended, but it is necessary to ensure previously that these three cooling powers are effectively provided by the TES-backed-up refrigeration cycle. The cooling power statics are obtained for all operating modes, where it is noticed that, as expected in the light of the complexity of the interconnection of both systems, the three relevant cooling powers are definitely not independent in most operating modes. Indeed, the statics give rise to combined power maps where the power limits of a given cooling power are expressed as non-linear functions of the remaining powers. This is another contribution of this work, since an accurate description of the power limits is key to addressing the scheduling strategy and ensuring the feasibility of the scheduled references. 

Another relevant contribution of this work is the development of the cooling power control, that is responsible for getting the TES-backed-up refrigeration system to provide the three required cooling powers, by driving the four manipulated variables available. However, the control problem turns out to be fully actuated, since it is also imperative to control or at least supervise the degree of superheating at the compressor intake to avoid mechanical breakdowns, thus another reference is imposed on this variable considering energy efficiency and operating constraints. In this article a decentralised strategy is proposed, where a decoupling matrix is calculated to reduce the high coupling between some of the controlled variables. Moreover, a low-level cascade strategy is applied to the control of the expansion valves, using the refrigeration mass flows that circulate through the evaporator and the TES tank as virtual manipulated variables.

The paper is organised as follows. Section \ref{secSystemDescription} describes concisely the TES tank and the existing refrigeration facility, as well as the interconnection between all elements. Section \ref{secSystemModelling} provides some details regarding the modelling of the refrigeration cycle and the TES tank separately, while a detailed combined model describing both the fast dynamics related to the refrigeration cycle and the slow dynamics due to the TES tank is presented. The cooling power control strategy is proposed in Section \ref{secControlStrategy}, where the different operating modes are also defined and the analysis of the existing coupling between the main cooling powers is performed. A thorough closed-loop simulation is presented, where the references on the three main cooling powers are imposed in such a way that all operating modes are tested. Finally, Section \ref{secConclusions} summarises the main conclusions to be derived, whereas some future work is proposed. 


\section{System description} \label{secSystemDescription}

\subsection{Notation}

Table \ref{tabSymbols} provides the symbols and subscript/superscript notation followed throughout the article. All fluid thermodynamic properties are rigorously considered in the simulation models, provided by the \emph{CoolProp} tool \cite{CoolProp}.

\begin{table}[h]
	\centering
	\caption{List of symbols and subscript/superscript notation}
	\label{tabSymbols}
	\scalebox{0.66}[0.66]{ \tabulinesep=0.5mm
		\begin{tabu} { L{1.3cm} L{8cm} C{1.1cm} | L{1.8cm} L{6cm} }
			\midrule
			\multicolumn{3}{c|}{\textbf{Italic symbols}} & \multicolumn{2}{c}{\textbf{Subscripts}} \\
			\midrule
			\emph{Symbol} & \emph{Description} & \emph{Units} & \emph{Symbol} & \emph{Description} \\ 
			\midrule
			$A$ & Opening & \% & $C$ & closed-loop \\
			$\bm{C}$ & Controller transfer matrix & -- & $c$ & condenser \\
			$\bm{\hat{C}}$ & Partial controller transfer matrix & -- & $comp$ & compressor \\
			$\bm{D}$ & Decoupling matrix & -- & $dp$ & dominant pole \\
			$\bm{\hat{D}}$ & Partial decoupling matrix & -- & $e$ & evaporator \\
			$\mathbold{f}$ & Non-linear function describing the condenser simplified model & -- & $in$ & inlet/input \\
			$\bm{H}$ & Linear model transfer matrix & -- & $int$ & intermediate fluid \\
			$\bm{\hat{H}}$ & Partial linear model transfer matrix & -- & $it$ & iterative \\
			$h$ & Specific enthalpy & J kg\textsuperscript{-1} & $lay$ & cylindrical layer \\
			$\bm{K}$ & Static gain matrix & -- & $out$ & outlet/output \\	
			$\dot m$ & Mass flow rate & g s\textsuperscript{-1} & $p$ & proportional \\
			$N$ & Compressor speed & Hz & $pcm$ & Phase Change Material \\
			$n$ & Number of elements (e.g. PCM cylinders) & -- & $SH$ & superheating \\	
			$P$ & Pressure & Pa & $sec$ & secondary fluid \\
			$\dot Q$ & Cooling power & W & $surr$ & surroundings \\
			$q$ & Vapour quality & -- & $TES$ & Thermal Energy Storage \\
			$s$ & Laplace variable & -- & $v$ & expansion valve \\
			$T$ & Temperature & K & $z$ & zero \\
			\cmidrule{4-5}
			$T_i$ & Integral time & s & \multicolumn{2}{c}{\textbf{Superscripts}} \\
			\cmidrule{4-5}
			$t$ & Time & min & \emph{Symbol} & \emph{Description} \\
			\cmidrule{4-5} 
			$U$ & Internal energy & J & $diag$ & diagonal \\
			$\mathbold{u}$ & Manipulated input vector & -- & $lat$ & latent state \\
			$\dot W$ & Mechanical power & W & $lat+$ & Maximum enthalpy latency point \\ 
			$\mathbold{w}$ & Input vector & -- & $lat-$ & Minimum enthalpy latency point \\
			$\mathbold{x}$ & State vector & -- & $max$ & maximum \\
			$\mathbold{y}$ & Output vector & -- & $min$ & minimum \\
			$\mathbold{Z}$ & Coefficient matrix & -- & $ref$ & reference \\
			\cmidrule{1-3}  
			\multicolumn{3}{c|}{\textbf{Greek symbols}} &  & \\
			\cmidrule{1-3}
			\emph{Symbol} & \emph{Description} & \emph{Units} &  & \\
			\cmidrule{1-3}
			$\gamma$ & \emph{Charge ratio} & -- &  & \\
			$\bm{\lambda}$ & Relative gain matrix & -- & & \\
			$\tau$ & Time constant & s & & \\
			$\mathbold{\psi}$ & Iterative error & -- & & \\
			\bottomrule
	\end{tabu}}
\end{table}

\subsection{Layout of the TES-backed-up refrigeration facility}

\begin{figure}[htbp]
	\centerline{\includegraphics[width=7cm,trim = 115 55 150 85,clip]
		{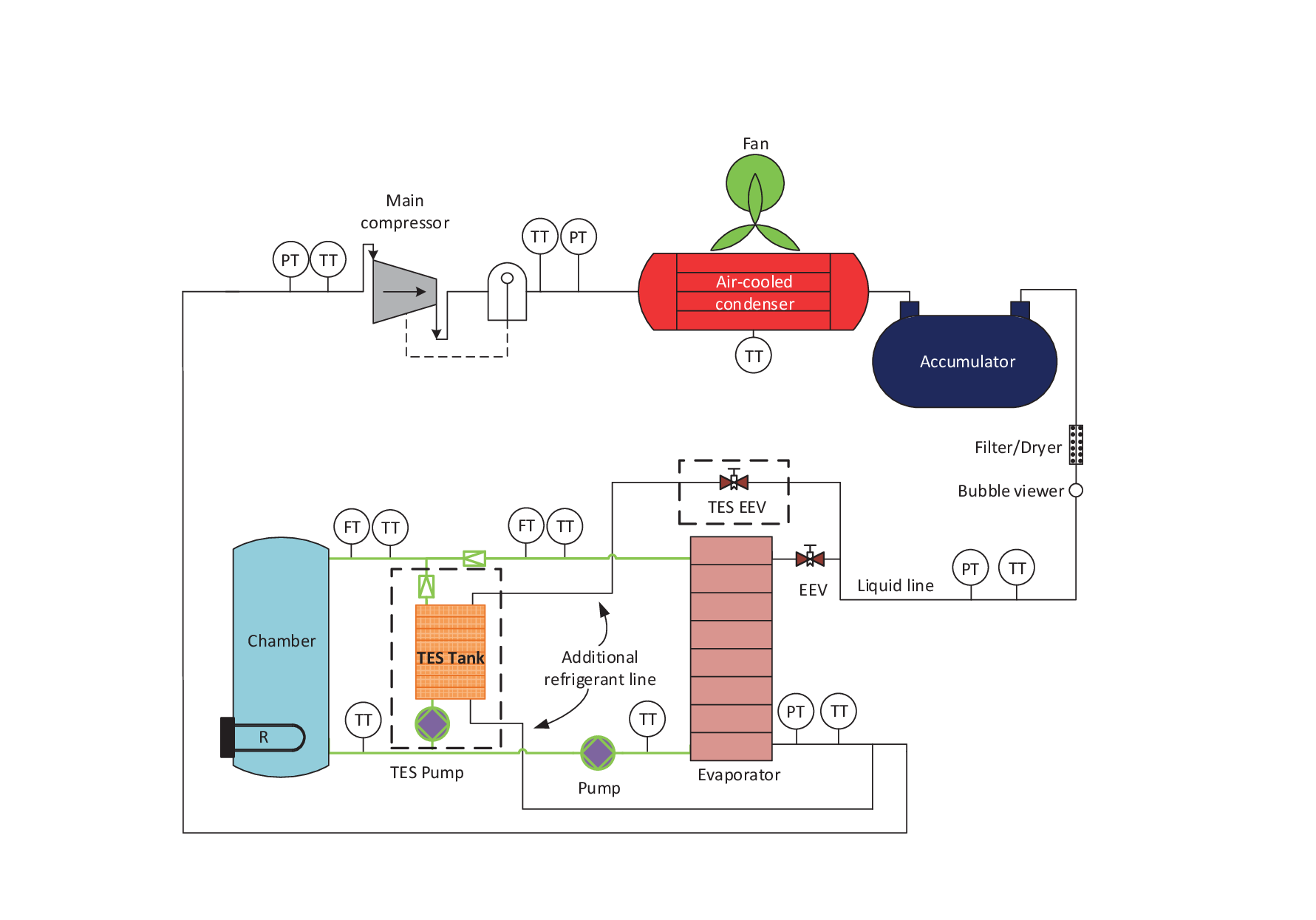}}
	\caption{Schematic diagram of the experimental refrigeration plant.}
	\label{figEsquemaPlanta}
\end{figure}

The plant under study is a two-compression-stage, two-load-demand experimental refrigeration facility located at the Department of Systems Engineering and Automatic Control at the University of Seville (Spain). Figure \ref{figEsquemaPlanta} shows a diagram of just the section corresponding to a one-compression-stage, one-load-demand refrigeration cycle, in which a PCM-based TES tank is intended to be included, along with other auxiliary elements such an additional expansion valve (all of them have been either highlighted or marked with a dash line in Figure \ref{figEsquemaPlanta}). The TES tank is the main element in the upgrading process undertaken on this facility, whose constructive details can be found in previous works by Bejarano \emph{et al.} \cite{GB_JE_2015,bejarano2015multivariable}. 

It is shown in Figure \ref{figEsquemaPlanta} that, as described in Section \ref{secIntroduction}, the TES tank is arranged in parallel with the evaporator, in such a way that two bundles of pipes run through it: one corresponding to the refrigerant (which must be deployed additionally to the original setup), devoted to charging the TES tank, and another one corresponding to the secondary fluid, the same circulating through the evaporator and pumped out from the cooled chamber, in this case devoted to discharging the TES tank. An additional pump (labelled as \emph{TES Pump} in Figures \ref{figEsquemaCicloPCM} and \ref{figEsquemaPlanta}) is necessary to drive the secondary fluid from the chamber to the TES tank, similar to that driving it to the evaporator. Technical details regarding the embedding of the TES tank in the existing refrigeration facility can be found in a previous work by the authors \cite{Bejarano2017NovelSchemePCM}. It is worth noting that this stage of the refrigeration system should satisfy a given cooling demand around $-20^{\circ}\mathrm{C}$, according to a typical reference level for freezing.

\begin{figure}[h]
	\centerline{\includegraphics[width=13cm,trim = 150 390 150 140,clip]
		{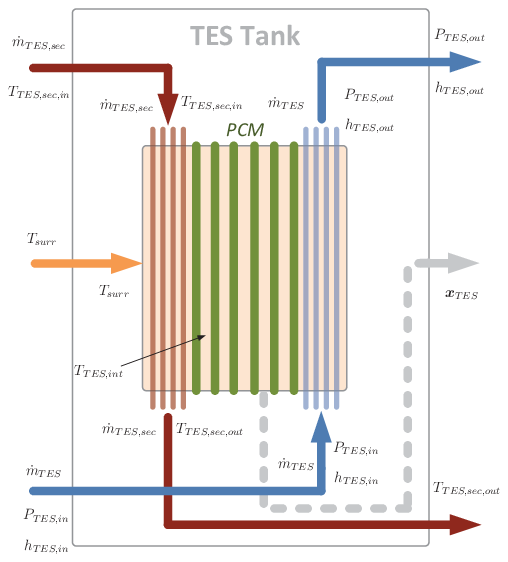}}
	\caption{Schematic picture of the proposed configuration of the TES tank and input-output conceptualisation.}
	\label{figTEStank_esquema}
\end{figure}

Figure \ref{figTEStank_esquema} shows a schematic picture of the proposed setup for the TES tank. As stated in Section \ref{secIntroduction}, the PCM is enclosed inside a number of steel cylinders. That represents the main difference between the actual design and that described in the previous work by the authors \cite{Bejarano2017NovelSchemePCM}, where the PCM was enclosed in spherical polymer capsules. The PCM encapsulation has been modified according to technical considerations, concerning the PCM filling and maintenance. The PCM cylinders are bathed in the so-called intermediate fluid, which is a fluid with high thermal conductivity and low heat capacity. Moreover, two bundles of pipes run through the tank, corresponding to the refrigerant (cold HTF) and the secondary fluid (warm HTF). All pipes are also dipped in the intermediate fluid. On the one hand, when charging the TES tank, this acts as an evaporator, since the refrigerant circulates through it, while evaporating and extracting heat from the intermediate fluid and then from the PCM enclosed in the cylinders. On the other hand, when discharging the TES tank, the secondary fluid circulates while transferring heat to the intermediate fluid and then to the PCM.

\begin{figure*}[h]
	\centering
	\subfigure[Plan view of the TES tank.]{
		\includegraphics[width=5.5cm,trim = 10 10 30 50,clip]{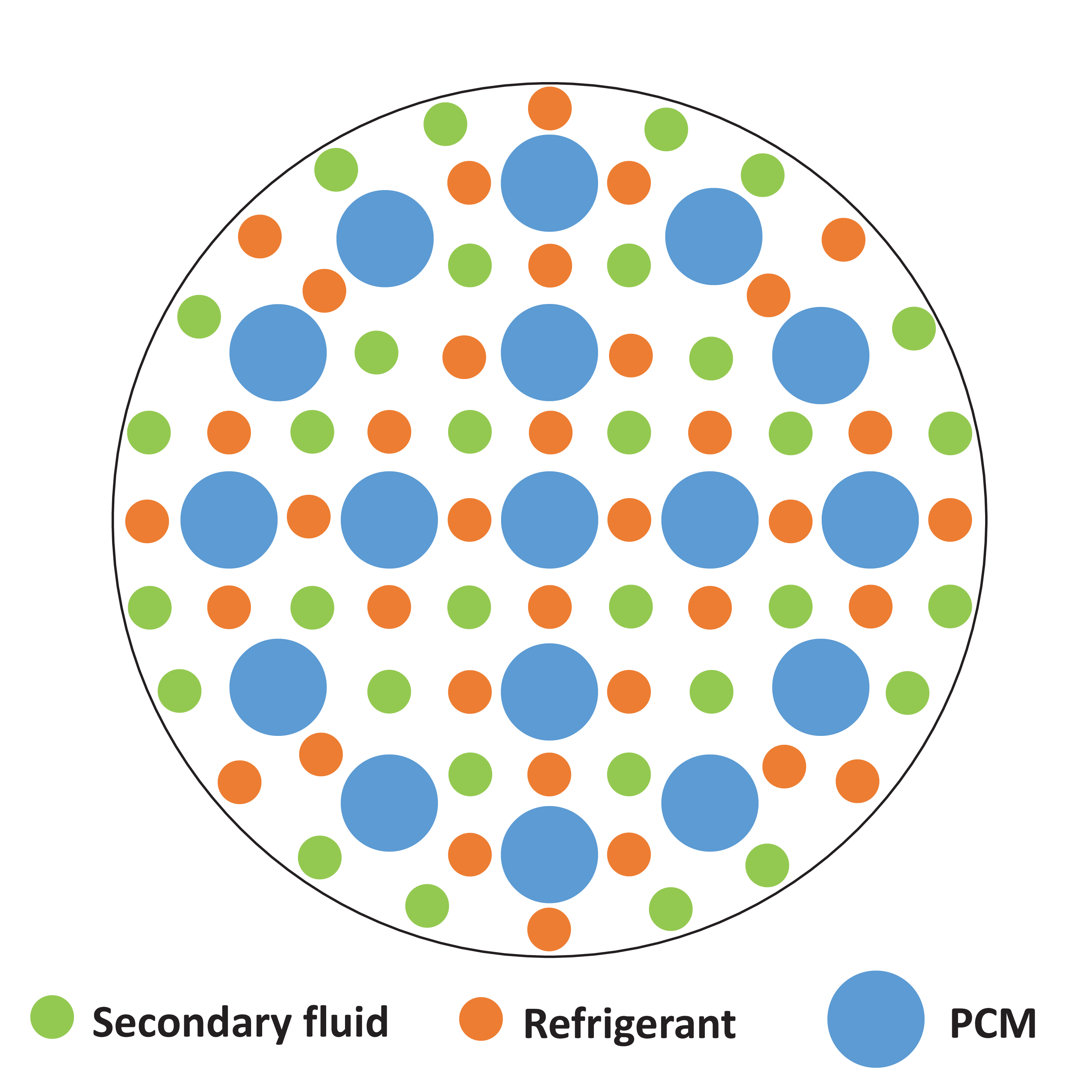}
		\label{figTESTank_real_a}
	}\subfigure[Picture taken during the constructive process of the TES tank.]{
		\includegraphics[width=7.5cm,trim = 0 0 0 0,clip]{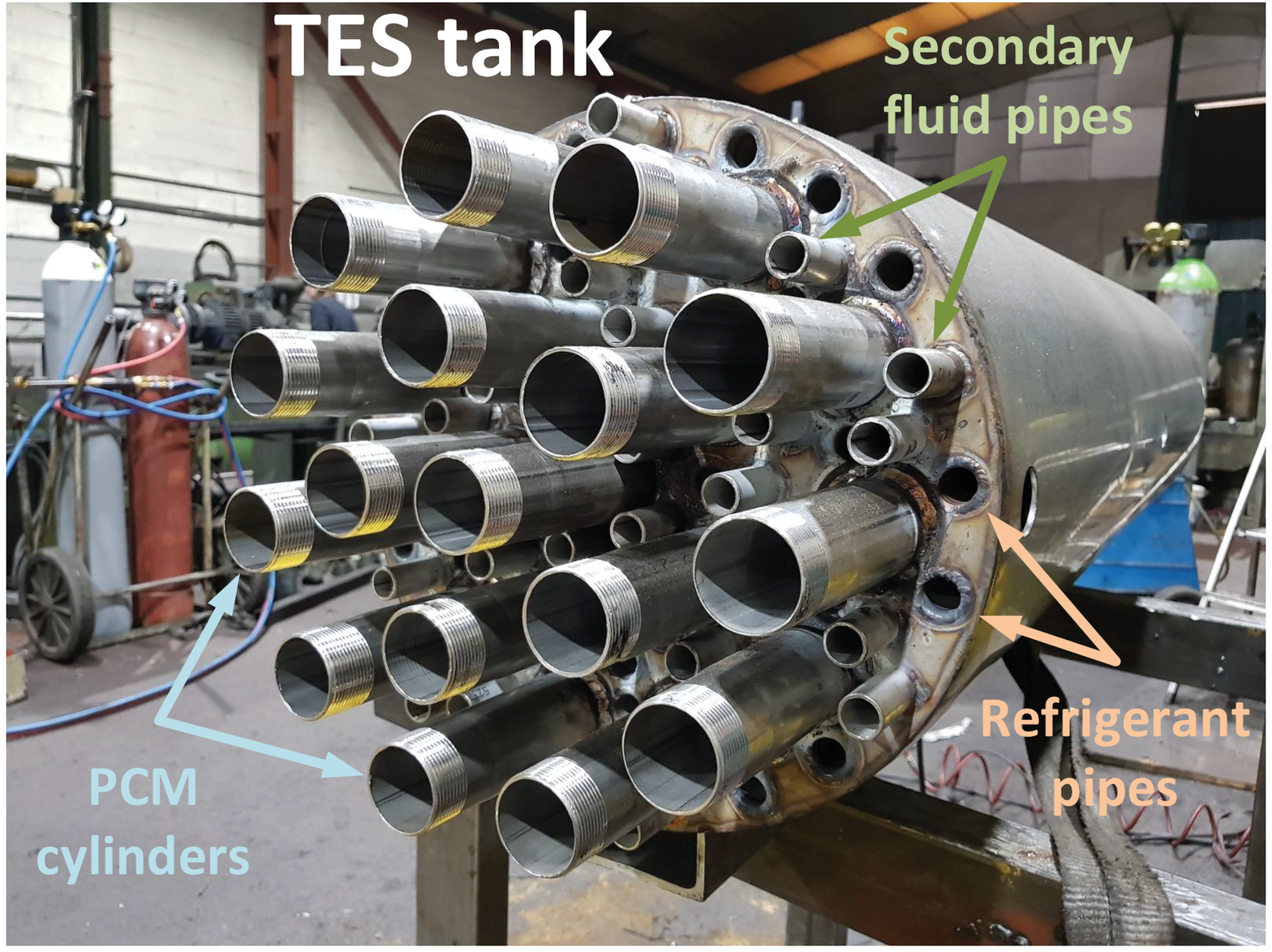}
		\label{figTESTank_real_b}
	}
	\caption{Plan schematic view and picture showing the actual distribution of pipes and PCM cylinders within the TES tank.}
	\label{figTESTank_real}
\end{figure*}

It is remarked that Figure \ref{figTEStank_esquema} is just a conceptual scheme. From a constructive point of view, all pipes and PCM cylinders are distributed in such a way that homogeneous heat transfer is achieved, as shown in Figure \ref{figTESTank_real}, where a picture taken during the constructive process of the TES tank (still under development) is also included.

Conceptual input-output representation is also included in Figure \ref{figTEStank_esquema}. The main input variables are the refrigerant mass flow rate, $\dot m_{TES}$, and the secondary fluid mass flow rate, $\dot m_{TES,sec}$, that essentially represent the extensive manipulated inputs to control the charging and discharging processes, respectively. Other important inputs are the inlet temperature of the secondary fluid, $T_{TES,sec,in}$, the inlet pressure and the specific enthalpy of the refrigerant, $P_{TES,in}$ and $h_{TES,in}$, and the ambient temperature outside the TES tank, $T_{surr}$. The pressure-specific enthalpy pair \{$P-h$\} has been selected to describe the thermodynamic state of the refrigerant at the TES tank inlet. The temperature-specific enthalpy pair  \{$T-h$\} could have also been chosen, instead, given that the refrigerant comes from the TES expansion valve and it is expected to be two-phase.

Concerning the output variables, the thermodynamic state of both the refrigerant and the secondary fluid is described by $T_{TES,sec,out}$ and the \{$P_{TES,out} - h_{TES,out}$\} pair, respectively. Moreover, the temperature of the intermediate fluid, $T_{TES,int}$, represents a state variable included in the state vector $\mathbold{x}_{TES}$, as well as a quantification of the specific enthalpy of the PCM. This quantification has been formulated by means of the cold-thermal energy storage ratio $\gamma_{TES}$, that can be obtained from the instantaneous enthalpy distribution inside the PCM cylinder. It will be referred as the \emph{charge ratio} for short, from now on, and it will be described in detail in Section \ref{secSystemModelling}.


\section{Modelling} \label{secSystemModelling}

\subsection{Refrigeration cycle} \label{subSecRefrigerationCycle}

Some recent works by the authors have addressed modelling and identification of the existing refrigeration facility, where every element has been identified separately in steady state \cite{bejarano2016identifying,rodriguez2017parameter}. In particular, regarding the heat exchangers, it has been shown that their heat-transfer-related parameters affect system statics but barely influence system dynamics. An identification procedure based on non-measurable refrigerant phase-change zones has been applied, considering an overall heat transfer coefficient at every zone. Consistent values of all parameters have been obtained considering only steady-state experimental data and some orders of magnitude found in the literature. The identified parameters have been validated considering different plant configurations.

Concerning dynamic modelling of the heat exchangers, that turn out to exhibit the dominant dynamics in the cycle, the \emph{switched moving boundary} (SMB) approach \cite{McKinley,BINLI} was found to achieve the best trade-off between accuracy and computational cost. Despite featuring lower complexity and computational load than previous formulations \cite{pangborn2015comparison}, the order of the SMB model is still high enough to restrict its application to simulation studies and even preventing the integration of the SMB approach in model-based control strategies. A simplified control-oriented model of the whole one-compression-stage, one-load-demand cycle has been proposed in a recent work by Bejarano \emph{et al.} \cite{Bejarano2017}, where some assumptions have been made to reduce the order of the original SMB model. In essence, the state vector of the whole cycle is reduced to that describing the slowest, and thus dominant, condenser dynamics, while the remaining elements in the cycle are statically modelled. Deviations between the simplified model and the original SMB model have been shown to be negligible with respect to dominant dynamics, whereas the computation time is drastically reduced. In accordance to this, the steady-state models of the compressor, expansion valve, and evaporator, along with the simplified dynamic model of the condenser already developed in the aforementioned works, will be also used when modelling the more complex cycle at hand, when the TES tank is added.    

\subsection{TES tank} \label{subSecTESTank}

Bejarano \emph{et al.} have recently studied the dynamic modelling of a very similar setup to that presented in this paper, where the only relevant difference is related to the PCM encapsulation, being in that work spherical capsules rather than current cylinders \cite{Bejarano2017NovelSchemePCM}. The two main approaches that can be found in the literature about modelling of PCM-based systems have been explored in that paper: a first-principle model and a finite-element one. 

On the one hand, the first approach (denoted as \emph{continuous model} in the paper by Bejarano \emph{et al.} \cite{Bejarano2017NovelSchemePCM}) accurately represents the dominant dynamics with minor computational load and it may be adapted to the current cylindrical geometry with minor changes. However, the \emph{continuous model} studies the evolution of a single inward freezing/melting front within the PCM element. It implies that only full charging/discharging operations can be simulated, since any series of partial charging/discharging processes would involve a potentially infinite-dimensional state vector. Recall that this model is intended to be integrated within a more complex modelling setup including the refrigeration cycle components, which, in turn, will be used to design efficient energy-management strategies. As a consequence, it is very likely that partial PCM charging/discharging processes have to be scheduled, according to economic and energy efficiency criteria. Therefore, this intrinsic limitation of the \emph{continuous model} becomes an excluding shortcoming.

On the other hand, the finite-volume model (denoted as \emph{discrete model} in the paper by Bejarano \emph{et al.} \cite{Bejarano2017NovelSchemePCM}) provides an accurate description of the dominant dynamics. In this approach, the PCM elements are conceptually divided into a certain number of layers $n_{lay}$, whose quantity represents a trade-off between accuracy and computational load. Unfortunately, this \emph{discrete model} is still computationally demanding, in fact, unacceptable given the intended use of the combined model. Then, a highly time-efficient version of the \emph{discrete model} has been recently presented, reducing drastically the simulation time and just incurring in affordable inaccuracies when describing the system behaviour \cite{bejarano2018efficient}. The discretisation time is increased, being the time period under consideration divided into some intervals when cold energy is assumed to be transferred at a constant rate. This time-efficient model is also the final choice in the work presented here.

\begin{figure}[htbp]
	\centerline{\includegraphics[width=7cm,trim = 225 325 90 235,clip]
		{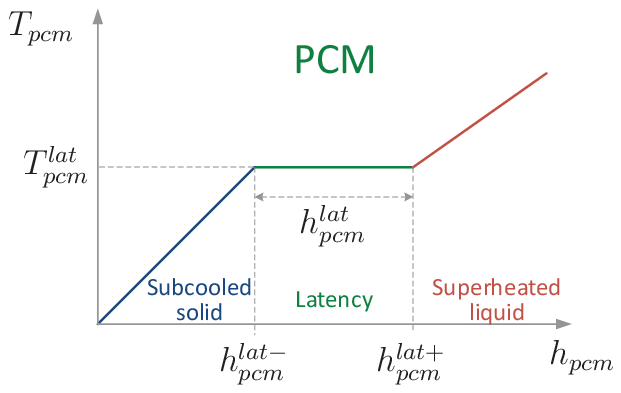}}
	\caption{Temperature-specific enthalpy diagram of the PCM.}
	\label{figThDiagrams}
\end{figure}

Regarding the estimation of the \emph{charge ratio} $\gamma_{TES}$, the PCM may be in solid phase (subcooled solid), liquid phase (superheated liquid), or in transition between both phases (latency). Figure \ref{figThDiagrams} represents the temperature-specific enthalpy diagram of the PCM, where $h_{pcm}^{lat-}$ just represents a reference level on the PCM specific enthalpy $h_{pcm}$, in such a way that $h_{pcm}^{lat+} \,=\, h_{pcm}^{lat-} \,+\, h_{pcm}^{lat}$.

The \emph{charge ratio} $\gamma_{TES}$ describes the charge state of the TES tank as a fraction of the maximum efficient cold-energy storage which the PCM is able to take on, provided that it remains in the latent zone. Heat transfer between the  intermediate fluid and the PCM is widely known to be more efficient if the latter remains in latent zone, and therefore it is desirable that the PCM specific enthalpy remains within the interval $[h_{pcm}^{lat-}, \ h_{pcm}^{lat+}]$. The \emph{maximum ''efficient'' cold-energy storage} $U_{TES}^{min}$ is defined as the thermal energy stored when the whole PCM volume reaches $h_{pcm}^{lat-}$, whereas the \emph{minimum ''efficient'' cold-energy storage} $U_{TES}^{max}$ is defined as the thermal energy stored when the whole PCM volume reaches $h_{pcm}^{lat+}$. According to these definitions, the \emph{charge ratio} $\gamma_{TES}$ is defined as a normalised index between 0 and 1, as indicated in Equation \eqref{eq_ChargeRatioDefinition}. It is remarked that, as $U_{TES}$ represents thermal energy, it is minimum when the maximum possible cold energy is stored in latent zone. 

\begin{equation}
	0 \leqslant \gamma_{TES} \,=\, \frac{U_{TES}^{max} - U_{TES}}{U_{TES}^{max} - U_{TES}^{min}} \leqslant 1 
	\label{eq_ChargeRatioDefinition}
\end{equation}

According to the uniform distribution of the PCM cylinders shown in Figure \ref{figTESTank_real} and the high thermal conductivity of the intermediate fluid, it is assumed that all PCM cylinders evolve in the same way. Hence, the overall stored thermal energy $U_{TES}$ can be supposed to be merely proportional to that stored inside every PCM cylinder $U_{pcm}$, and thus the \emph{charge ratio} $\gamma_{TES}$ can be computed as indicated in Equation Set \eqref{eq_ChargeRatioDefinition2}, where $n_{pcm}$ refers to the number of PCM cylinders within the TES tank.

\begin{equation}
	\begin{aligned}
	U_{TES} \,&=\, n_{pcm}\,U_{pcm}\; \\
	U_{TES}^{min} \,&=\, n_{pcm}\,U_{pcm}^{min}\; \\
	U_{TES}^{max} \,&=\, n_{pcm}\,U_{pcm}^{max} \\
	\gamma_{TES} \,&=\, 
	\frac{U_{pcm}^{max} - U_{pcm}}
	{U_{pcm}^{max} - U_{pcm}^{min}} \\
	\end{aligned}
	\label{eq_ChargeRatioDefinition2}
\end{equation}

\begin{figure}[h]
	\centerline{\includegraphics[width=10cm,trim = 150 450 185 150,clip]
		{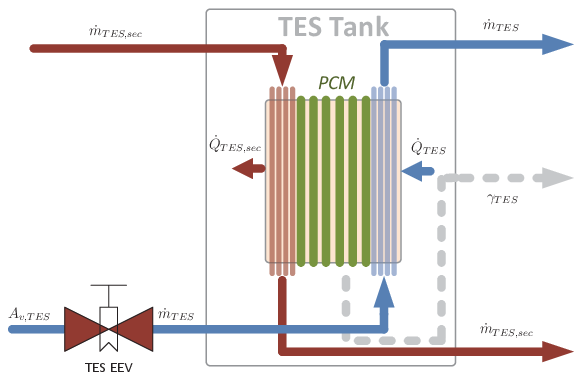}}
	\caption{Simulation setup including the TES tank and the corresponding expansion valve.}
	\label{figTES_and_Av}
\end{figure}

\begin{figure}[h]
	\centering
	\subfigure[\emph{Charge ratio} for diverse charging processes.]{
		\includegraphics[width=7cm] {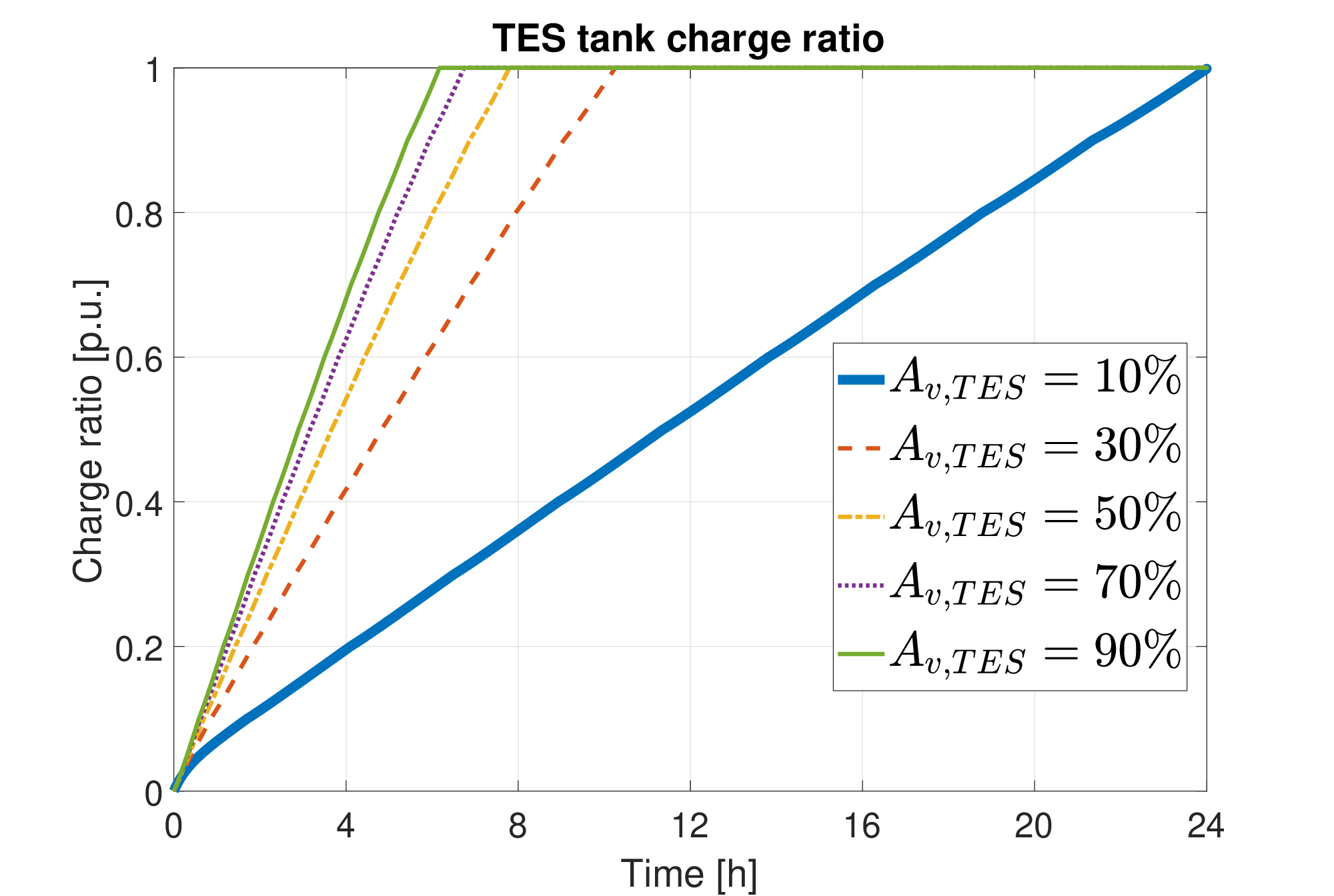}
	}\subfigure[Relationship between $\dot{Q}_{TES}$ and $\gamma_{TES}$ for different charging processes.]{
		\includegraphics[width=7cm] {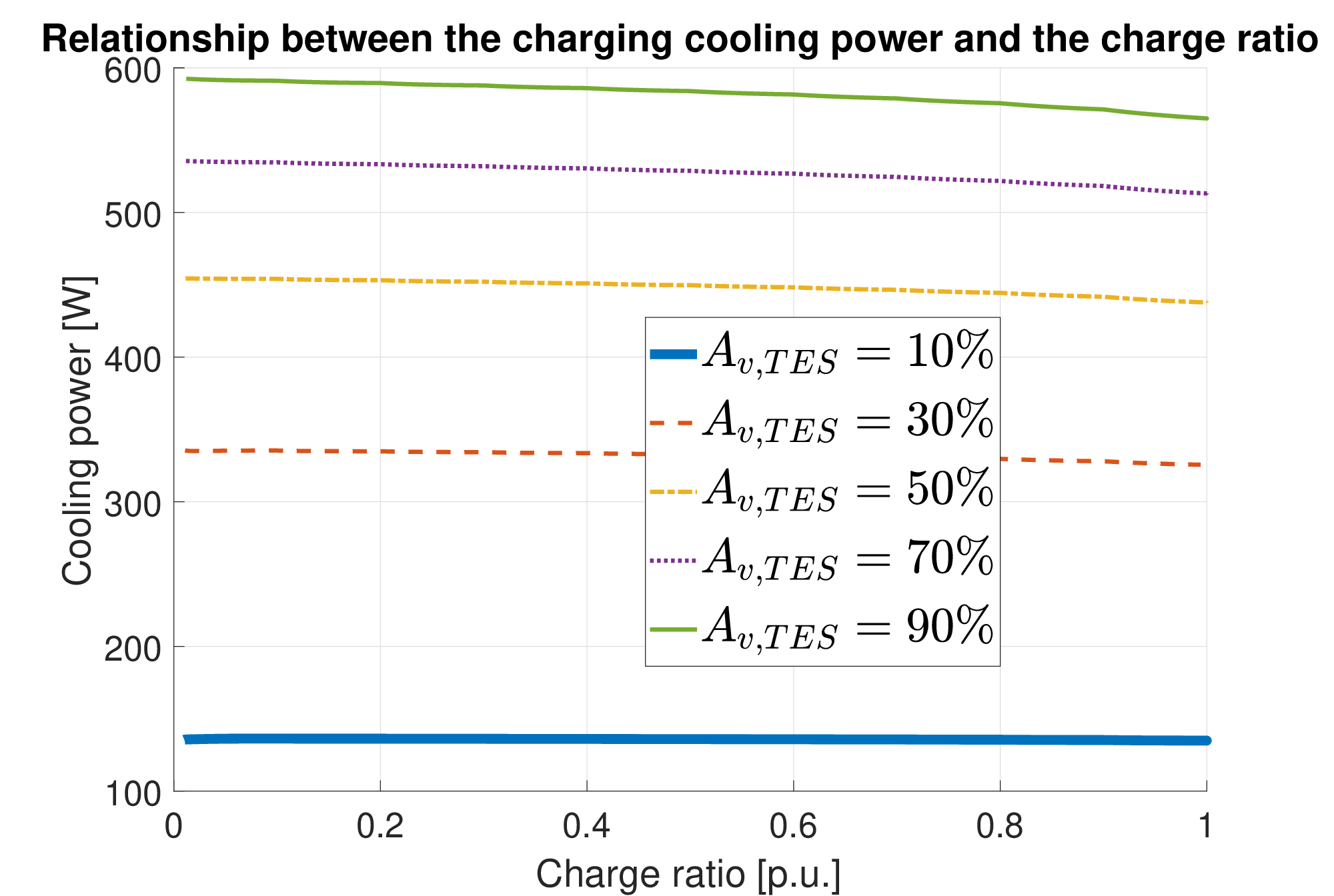}
	}
	\subfigure[\emph{Charge ratio} for diverse discharging processes.]{
		\includegraphics[width=7cm] {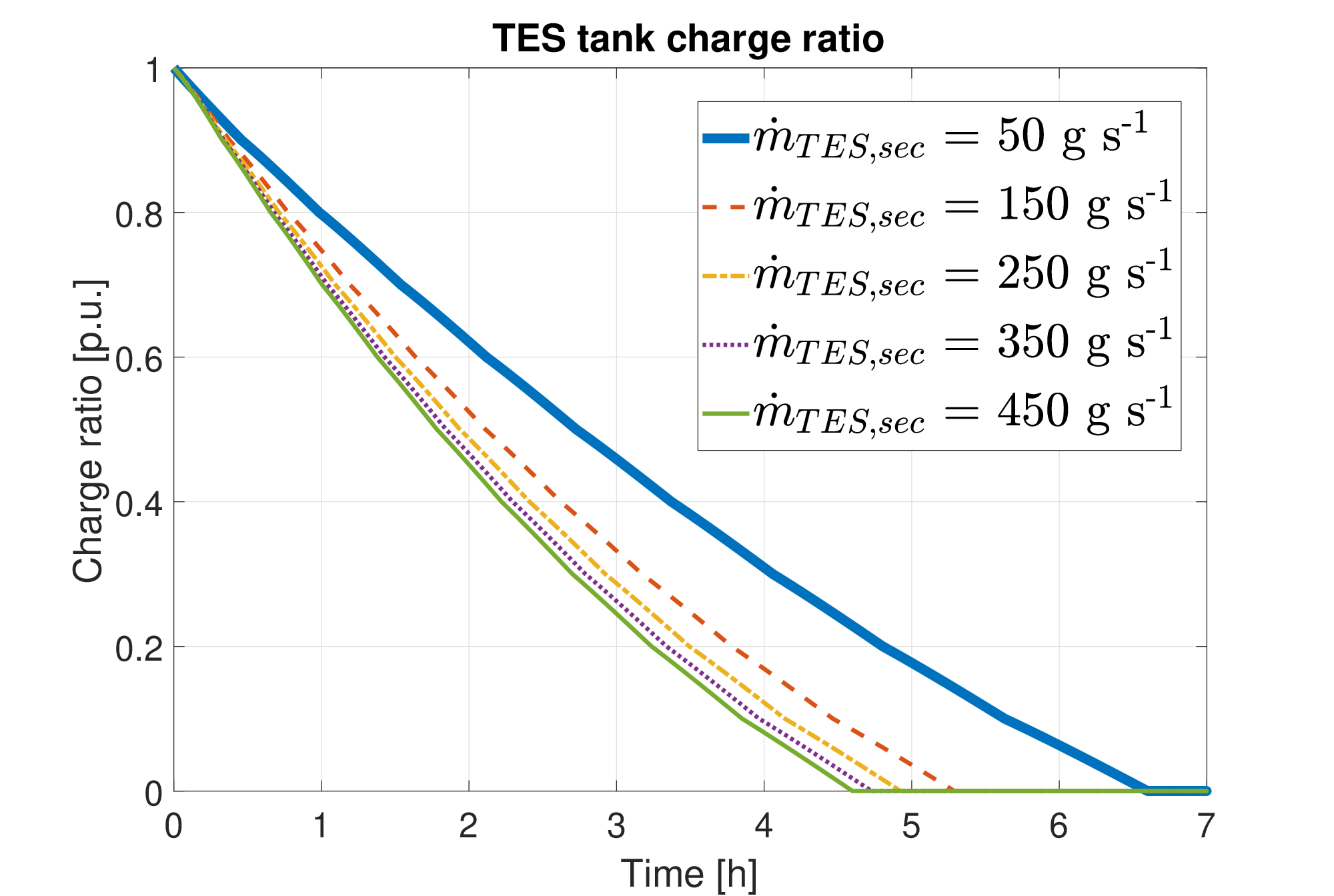}	
	}\subfigure[Relationship between $\dot{Q}_{TES,sec}$ and $\gamma_{TES}$ for different discharging processes.]{
		\includegraphics[width=7cm] {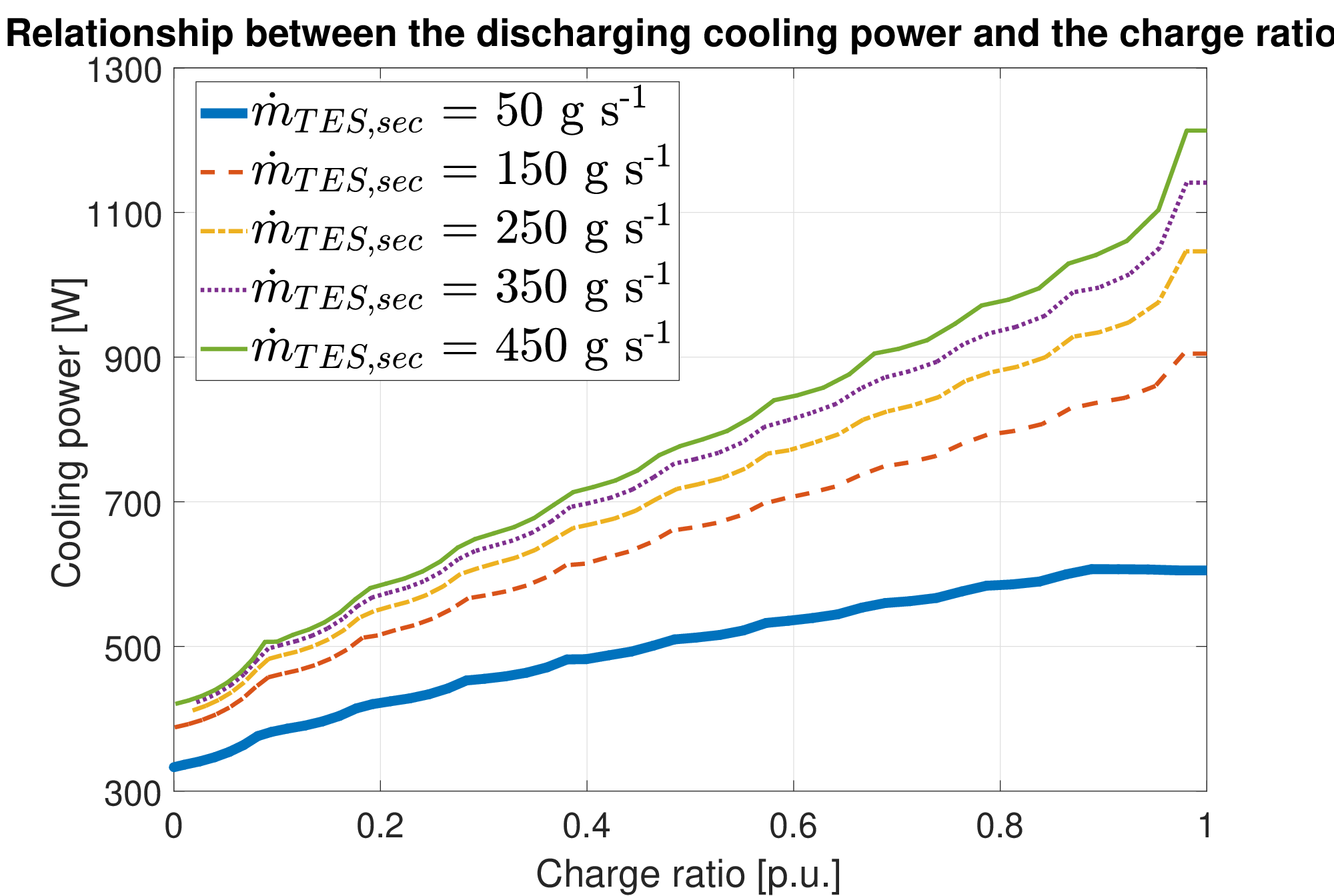}
		
	}		
	\caption{TES tank behaviour for different charging/discharging processes.}
	\label{figTEScharging_discharging}				
\end{figure}

Eventually, the behaviour of the TES tank regarding the \emph{charge ratio} $\gamma_{TES}$ is analysed for complete charging/discharging processes, as well as the variation of the cooling powers while $\gamma_{TES}$ evolves. The setup represented in Figure \ref{figTES_and_Av} has been applied to obtain the simulation results presented in Figure \ref{figTEScharging_discharging}, where the manipulated inputs turn out to be the TES expansion valve opening $A_{v,TES}$, since the latter has been externally added to the TES tank model, and the secondary mass flow $\dot{m}_{TES,sec}$. In fact, regarding the discharging process, the actual manipulated variable would be the TES pump speed/power; anyway, given a desired feasible secondary mass flow $\dot{m}_{TES,sec}$, a simple low-level controller could drive the TES pump to impulse the desired secondary mass flow. The remaining inputs to the TES tank model shown in Figure \ref{figTEStank_esquema} remain constant, being their values those indicated in Table \ref{tabInputTES}. 

\begin{table}[htbp]
	\centering
	\caption{TES tank input values}
	\label{tabInputTES}
	\scalebox{0.75}[0.75]{ \tabulinesep=0.5mm
		\begin{tabu} { L{8.0cm} C{2.9cm} C{4cm} C{1.2cm} }
			\toprule
			\emph{\textbf{Variable}} & \emph{\textbf{Symbol}} & \emph{\textbf{Value}} & \emph{\textbf{Unit}} \\ 
			\toprule
			\multicolumn{4}{c}{\textbf{Charge}}\\
			\bottomrule	
			TES expansion valve opening & $A_{v,TES}$ & 10/30/50/70/90 &\% \\
			Inlet pressure of the refrigerant & $P_{TES,in}$ & 10\textsuperscript{5} &Pa \\
			Inlet specific enthalpy of the refrigerant & $h_{TES,in}$ & 258 $\cdot$ 10\textsuperscript{3} & J kg\textsuperscript{-1}\\
			Initial temperature of the intermediate fluid & $T_{TES,int} (t=0)$ & 242.15 & K \\
			Surroundings temperature& $T_{surr}$ & 293.15 & K \\
			\toprule
			\multicolumn{4}{c}{\textbf{Discharge}}\\
			\bottomrule	
			Secondary mass flow & $\dot{m}_{TES,sec}$ & 50/150/250/350/450 & g s\textsuperscript{-1} \\
			Inlet temperature of the secondary fluid  & $T_{TES,sec,in}$ & 253.15 & K\\
			Initial temperature of the intermediate fluid & $T_{TES,int} (t=0)$ & 246.15 & K \\
			Surroundings temperature& $T_{surr}$ & 293.15 & K \\
			\bottomrule			
	\end{tabu}}
\end{table}

It can be observed in Figure \ref{figTEScharging_discharging} that, as expected, the greater the charging/discharging cooling power, the shorter the time for the TES tank to be fully charged/discharged. However, the relationship between cooling power and total charging/discharging time is highly non-linear. Moreover, it can be noticed that the cooling power transferred decreases as $\gamma_{TES}$ evolves, both during the charging and the discharging processes. This is due to the extra thermal resistance caused by the inward growing freezing/melting shell in sensible zone of the PCM cylinders, being this effect much more significant in the case of the discharging process. Anyway, these simulation results should be qualitatively analysed, especially regarding the total charging/discharging times, since the assumption that some input variables shown in Table \ref{tabInputTES} remain constant (for instance, $P_{TES,in}$ and $h_{TES,in}$) is very strong. Indeed, when considering the whole refrigeration cycle those variables are more likely to be continuously varying according to the action of the remaining manipulated variables, as well as the self-adjustment of the cycle regarding pressures/enthalpies as the TES charges/discharges.  


\subsection{Interconnection and time scales} \label{subSecInterconnection}

In this subsection the interconnection of the separate models previously described is addressed to develop a joint dynamic model of the enhanced refrigeration cycle. This interconnection is not definitely trivial matter, since the TES tank is not merely added to the refrigeration cycle at its outlet, but inserted within the cycle, thus the topology is modified. Moreover, the time scales of both subsystems are very different, as detailed later, what increases the difficulty of the integral modelling by introducing some frequency issues.  

The heat-transfer-related dynamics of the TES tank are not comparable to those of the refrigeration cycle, since appropriate sampling times for the TES tank model are in the order of several minutes, due to the slow dynamics caused by the low heat capacity of the intermediate fluid, whereas the refrigeration cycle requires sampling times in the order of a few seconds, as analysed in the aforementioned literature \cite{bejarano2017suboptimal,bejarano2018efficient}. Therefore, two very different time scales arise: one linked to the fast dynamics of the refrigeration cycle, and another one slower, related to heat transfer within the TES tank.

Regarding the fast refrigeration cycle dynamics, the interconnection of the original cycle components with the TES tank is analysed. As observed in Figure \ref{figEsquemaCicloPCM}, from the point of view of the refrigerant, the TES tank acts as an evaporator when it is charged, since the refrigerant evaporates while removing heat from the intermediate fluid and then from the PCM cylinders, but this last heat transfer occurs at a much lower rate. Therefore, always considering the fast time scale, the refrigerant just transfers heat with the intermediate fluid, whose thermal dynamics are much slower than the fast ones which we focus right now on, which causes its temperature to be considered as constant. As a result, the TES tank does not add intrinsic fast dynamics to the cycle, and it may be modelled as a static element concerning the fast time scale, just like the evaporator. The same argument is applicable to the secondary fluid when the TES tank is discharged, since it also transfers heat to the intermediate fluid, not directly to the PCM cylinders.

As a consequence, the condenser continues to concentrate the dominant refrigeration cycle dynamics. A modelling structure of the whole cycle similar to that presented in Appendix A of the work by Bejarano \emph{et al.} for the canonical refrigeration cycle \cite{bejarano2017suboptimal}, is proposed, where the steady-state models of the evaporator, the compressor, both expansion valves, and the TES tank (regarding the fast time scale) are used to calculate the condenser boundary conditions at each instant, in particular the refrigerant mass flow $\dot{m}$ and inlet specific enthalpy $h_{c,in}$. Moreover, some interesting output variables such as the cooling power transferred to the secondary fluid at the evaporator $\dot{Q}_{e,sec}$, the charging cooling power $\dot{Q}_{TES}$, the discharging cooling power $\dot{Q}_{TES,sec}$, the degree of superheating at the compressor intake $T_{SH}$, and the mechanical power consumed by the compressor $\dot{W}_{comp}$ are also computed. Eventually, the TES tank state vector $\bm{x}_{TES}$ is also computed as an output of the TES tank model. An iterative procedure has been designed to solve the non-linear equation system generated by the steady-state models of the cycle components, which is graphically described in Figure \ref{figModeloEstatico_V_C_EE_TES}. This solution procedure is based on that presented in Appendix A of the work by Bejarano \emph{et al.} \cite{bejarano2017suboptimal} for the canonical cycle, while the steady-state models of the compressor, expansion valves, and evaporator are detailed in a previous work by Bejarano \emph{et al.} \cite{Bejarano2017}. 

\begin{figure*}[htbp]
	\centering
	\includegraphics[width=13cm,trim = 110 450 145 120,clip]
	{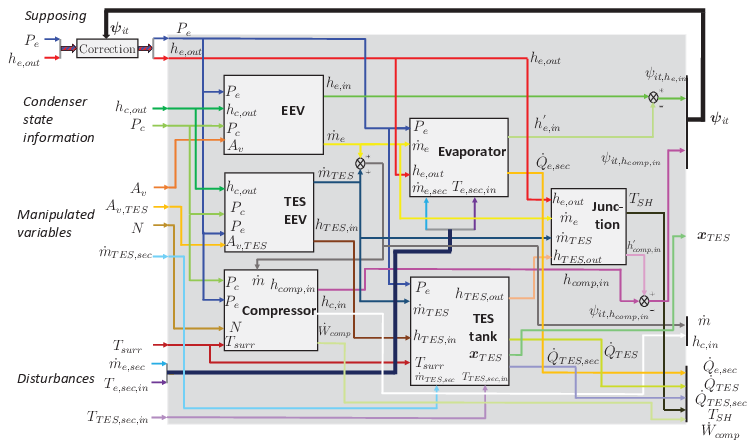}
	\caption{Iterative procedure to solve the non-linear equation system defined by the steady-state models of the fast components of the TES-backed-up refrigeration cycle.}
	\label{figModeloEstatico_V_C_EE_TES}
\end{figure*}

The block labelled as \emph{Junction} refers to the point where the refrigerant flows coming from the evaporator and the TES tank merge, indicated as point A in Figure \ref{figEsquemaCicloPCM}. The mass and power balance equations of the \emph{Junction} block are detailed in Equation Set \eqref{eq_JunctionBlock}, where the refrigerant-specific functions $T = T(P,q)$ and $T = T(P,h)$ are used to estimate the thermodynamic properties required, given by the \emph{CoolProp} tool \cite{CoolProp}. Note that $q$ refers to the vapour quality, in such a way that $q=1$ involves that the corresponding thermodynamic property is calculated considering saturated vapour. 
 
\begin{equation}
	\begin{aligned}
	h_{comp,in}^{'} &= \dfrac{\dot{m}_e \, h_{e,out} + \dot{m}_{TES} \, h_{TES,out}}{\dot{m}_e + \dot{m}_{TES}} \\
	T_e &= T(P=P_e,q=1) \\
	T_{comp,in} &= T(P=P_e,h=h_{comp,in}^{'}) \\
	T_{SH} & = T_{comp,in} - T_e \\
	\end{aligned}
	\label{eq_JunctionBlock}
\end{equation}

Concerning the block labelled as \emph{TES tank}, the equations modelling heat transfer within the TES tank between the intermediate fluid and the PCM cylinders, the refrigerant, and the secondary fluid described in the aforementioned modelling works by Bejarano \emph{et al.} \cite{Bejarano2017NovelSchemePCM,bejarano2018efficient} are included, while the time-efficient algorithm proposed in the latter work is applied to provide the evolution of the state vector $\mathbold{x}_{TES}$.   

Given the solution strategy for the steady-state part described in Figure \ref{figModeloEstatico_V_C_EE_TES}, it is integrated with the condenser dynamic model, as shown in Figure \ref{figDetailedModel}. The condenser dynamic equations are described in detail in Appendix A of the work by Bejarano \emph{et al.} \cite{bejarano2017suboptimal}. The detailed model of the TES-backed-up refrigeration cycle, focused on the fast dynamics caused by the refrigerant circulation, is one of the main contributions of the work. It must be integrated using a small sampling time (in the range of a few seconds), since it describes the refrigeration cycle dynamics in addition to the slow ones related to the intermediate fluid and its heat transfer with the PCM cylinders. Input, states, and outputs of this detailed model are also described  in Figure \ref{figDetailedModel}. 

\begin{figure}[htbp]
	\centering
	\includegraphics[width=14cm,trim = 130 460 110 140,clip]
	{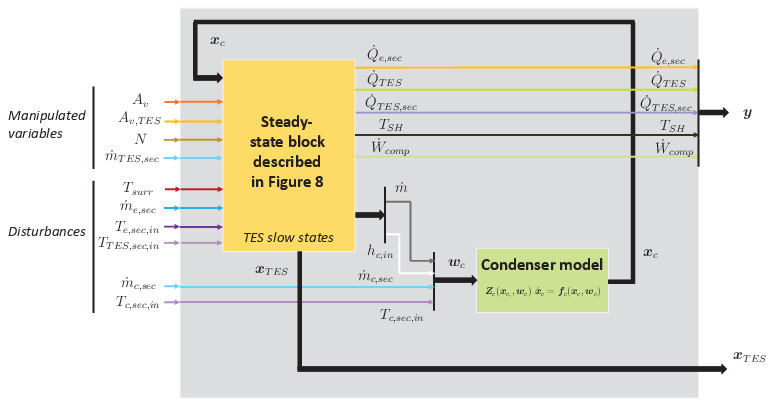}
	\caption{Detailed model of the TES-backed-up refrigeration cycle.}
	\label{figDetailedModel}
\end{figure}


\section{Cooling power control} \label{secControlStrategy}

\subsection{Overview} \label{subSecOverview}

\begin{figure}[htbp]
	\centering
	\includegraphics[width=17.0cm,trim = 220 295 5 170,clip]
	{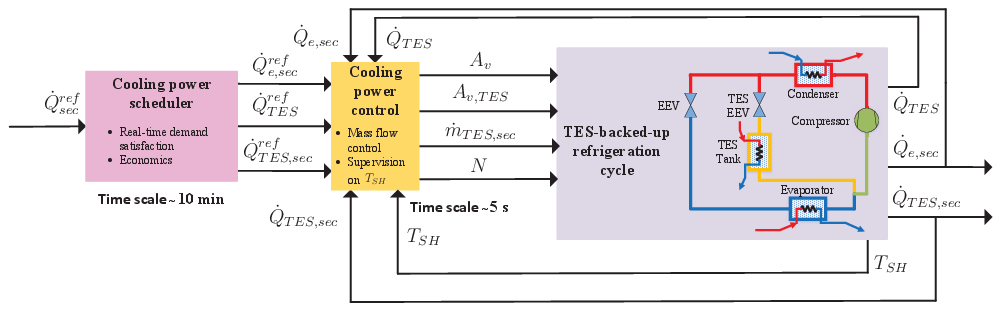}
	\caption{Scheduling and control strategy for the TES-backed-up refrigeration system.}
	\label{figControlStrategy}
\end{figure}

Figure \ref{figControlStrategy} represents the proposed scheduling and control strategy, already outlined in previous sections. Given a certain cooling demand to be provided to the secondary fluid, $\dot{Q}_{sec}^{ref}$, the top-level scheduler is intended to compute the references on the three cooling powers involved: $\dot{Q}_{e,sec}^{ref}$, $\dot{Q}_{TES}^{ref}$, and $\dot{Q}_{TES,sec}^{ref}$, in order to ensure the real-time demand satisfaction and considering economic criteria. The cooling power controller is then responsible for getting the TES-backed-up refrigeration system to provide the three required cooling powers, by manipulating the manipulated variables available. The scheduler is focused on the slower dynamics related to heat transfer within the TES tank, whose time scale is in the order of minutes, while the cooling power controller is focused on the faster dynamics related to the refrigerant circulation, whose time scale is in the order of a few seconds.

This work is focused on the cooling power control, that turns out to be a multivariable problem where, in the most general case, there are four manipulated variables available ($N$, $A_v$, $A_{v,TES}$, and $\dot{m}_{TES,sec}$) and four output variables to be controlled ($\dot{Q}_{e,sec}$, $\dot{Q}_{TES}$, $\dot{Q}_{TES,sec}$, and $T_{SH}$). As studied in previous works in the literature, refrigeration cycles are strongly non-linear systems with high coupling between variables \cite{Shen2010,rodriguez2018robust}. In this case, the addition of the TES tank to our setup increases complexity and coupling, specially regarding $\dot{Q}_{e,sec}$ and $\dot{Q}_{TES}$, whose respective refrigerant mass flows are connected at point A in Figure \ref{figEsquemaCicloPCM}. Moreover, the degree of superheating is measured at this point A, which matches the compressor intake, thus $T_{SH}$ is also very coupled with these two cooling powers. However, the interaction between $\dot{Q}_{TES,sec}$ and the TES-backed-up refrigeration cycle is expected to be minor, since it is another fluid circuit not tied to the compressor, and furthermore $\dot{Q}_{e,sec}$ and $\dot{Q}_{TES,sec}$ only interact at point B in Figure \ref{figEsquemaCicloPCM}, defining the inlet conditions of the secondary fluid at the refrigerated chamber.

\begin{figure}[htbp]
	\centering
	\includegraphics[width=13cm,trim = 120 310 45 160,clip]
	{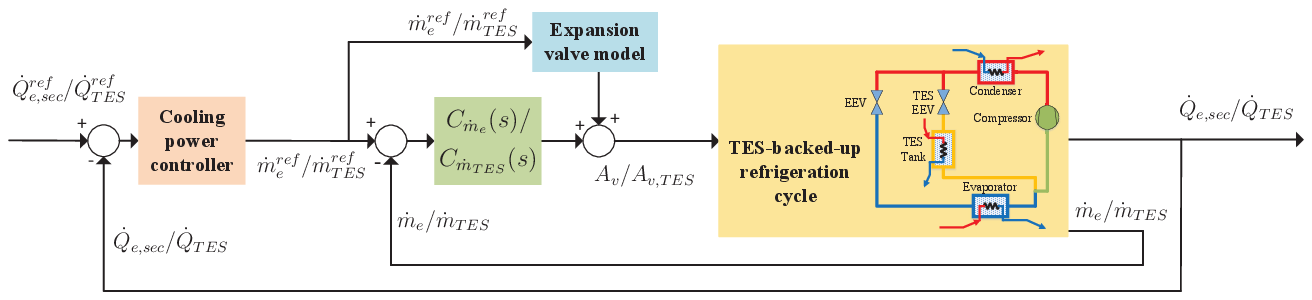}
	\caption{Cascade strategy applied to the expansion valve control.}
	\label{figExpansionValveControl}
\end{figure}

Firstly, in order to reduce the non-linear features added by the expansion valve models, the refrigerant mass flows $\dot{m}_e$ and $\dot{m}_{TES}$ are proposed to be used as virtual manipulated variables. The expansion valve model is then used as a feedforward contribution to the refrigerant mass flow low-level control, implemented as a PI controller $C_{\dot{m}_e}(s)$/$C_{\dot{m}_{TES}}(s)$, as indicated in Figure \ref{figExpansionValveControl}. This cascade strategy is applied for both expansion valves, and it is also expected to be applied for the TES pump, when its model is added to the simulator. In that case, $\dot{m}_{TES,sec}$ would be also considered as a virtual manipulated variable, while the TES pump model would be used as a feedforward contribution to the secondary mass flow low-level control. 

Secondly, regarding the control of the degree of superheating, a supervisory strategy is proposed, where the main objective is to hold the degree of superheating $T_{SH}$ over a minimum positive value ($T_{SH}^{min}$ = 2ºC has been imposed). To calculate the reference $T_{SH}^{ref}$, energy efficiency of the refrigeration cycle is considered. As widely known, energy efficiency is usually assessed in the context of refrigeration systems using the Coefficient of Performance ($COP$), which is defined as the ratio between the cooling power generated (in this case considering the evaporator and/or the TES tank charging process) and the mechanical power provided by the compressor. Previous studies show that high values of the $COP$ are achieved when the compressor speed $N$ is set as low as possible, provided that the cooling demand is satisfied \cite{Bejarano2017}. Therefore, the set point $T_{SH}^{ref}$ is calculated in such a way that the compressor speed $N$ matches the minimum value in steady state, provided that the cooling demand falls within the admissible range, but no $T_{SH}^{ref}$ under $T_{SH}^{min}$ is allowed. In other words, the compressor speed $N$ is forced to be as low as possible while holding $T_{SH}$ over $T_{SH}^{min}$. A PI controller $C_{T_{SH}}(s)$ is proposed to drive the compressor speed $N$ to get the degree of superheating to track $T_{SH}^{ref}$. More information about the $T_{SH}^{ref}$ calculation algorithm and its application to the original refrigeration facility can be found in the work by Bejarano \emph{et al.} \cite{bejarano2018optimization}.

\subsection{Operating modes} \label{subSecOperatingModes}

\begin{figure}[h]
	\centering
	\includegraphics[width=13cm,trim = 140 425 160 125,clip]
	{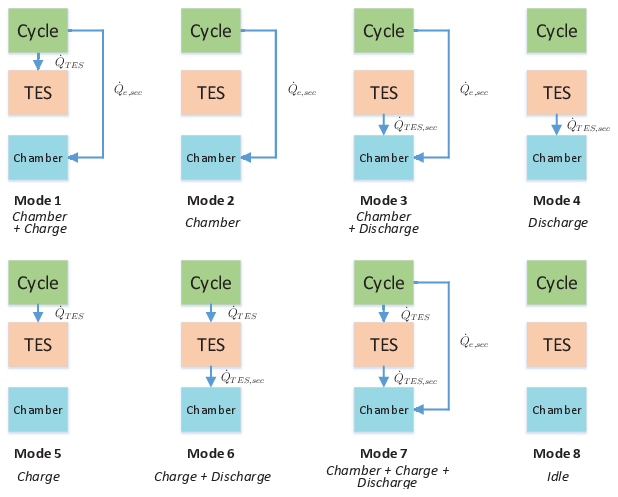}
	\caption{Operating modes of the TES-backed-up refrigeration cycle.}
	\label{figOperatingModes}
\end{figure}

A set of operating modes of the TES-backed-up refrigeration cycle has been defined, according to the different combinations of the activation state of three main cooling powers involved: the cooling power transferred to the secondary fluid at the evaporator $\dot{Q}_{e,sec}$, the charging cooling power $\dot{Q}_{TES}$, and the discharging cooling power $\dot{Q}_{TES,sec}$. The operating modes are described graphically in Figure \ref{figOperatingModes}.

Mode 1 is defined as that providing a certain cooling power $\dot{Q}_{e,sec}$ to the secondary fluid at the evaporator, while charging the TES tank with cooling power $\dot{Q}_{TES}$. Mode 2 is considered when the secondary fluid is cooled only at the evaporator, while the TES tank evolves only due to thermal losses. Mode 3 is defined as that providing cooling power to the secondary fluid both at the evaporator and the TES tank, which is discharged at a rate given by the cooling power $\dot{Q}_{TES,sec}$. Mode 4 is considered when the refrigeration cycle is stopped and there is no refrigerant circulating through the cycle, but the cooling demand is satisfied by discharging the TES tank. Mode 5 requires no cooling demand, since the refrigeration cycle is intended to charge exclusively the TES tank at a rate given by the charging cooling power $\dot{Q}_{TES}$. In mode 6, the refrigeration cycle charges the TES tank, but it is simultaneously discharged while the secondary fluid is cooled with power $\dot{Q}_{TES,sec}$. Mode 7 is the most complicated one, since all the cooling powers are involved: the refrigeration cycle cools the secondary fluid at the evaporator while charging the TES tank, which simultaneously provides additional cooling power to the secondary fluid. Eventually, mode 8 is the easiest, since all the cooling powers are zero, the refrigeration cycle is stopped, and the TES tank evolves only due to thermal losses (stand-by regime).

\subsection{Coupling analysis and power limits} \label{subSecCoupling}

\begin{figure}[h]
	\centering
	\includegraphics[width=13.5cm,trim = 0 0 0 0,clip] {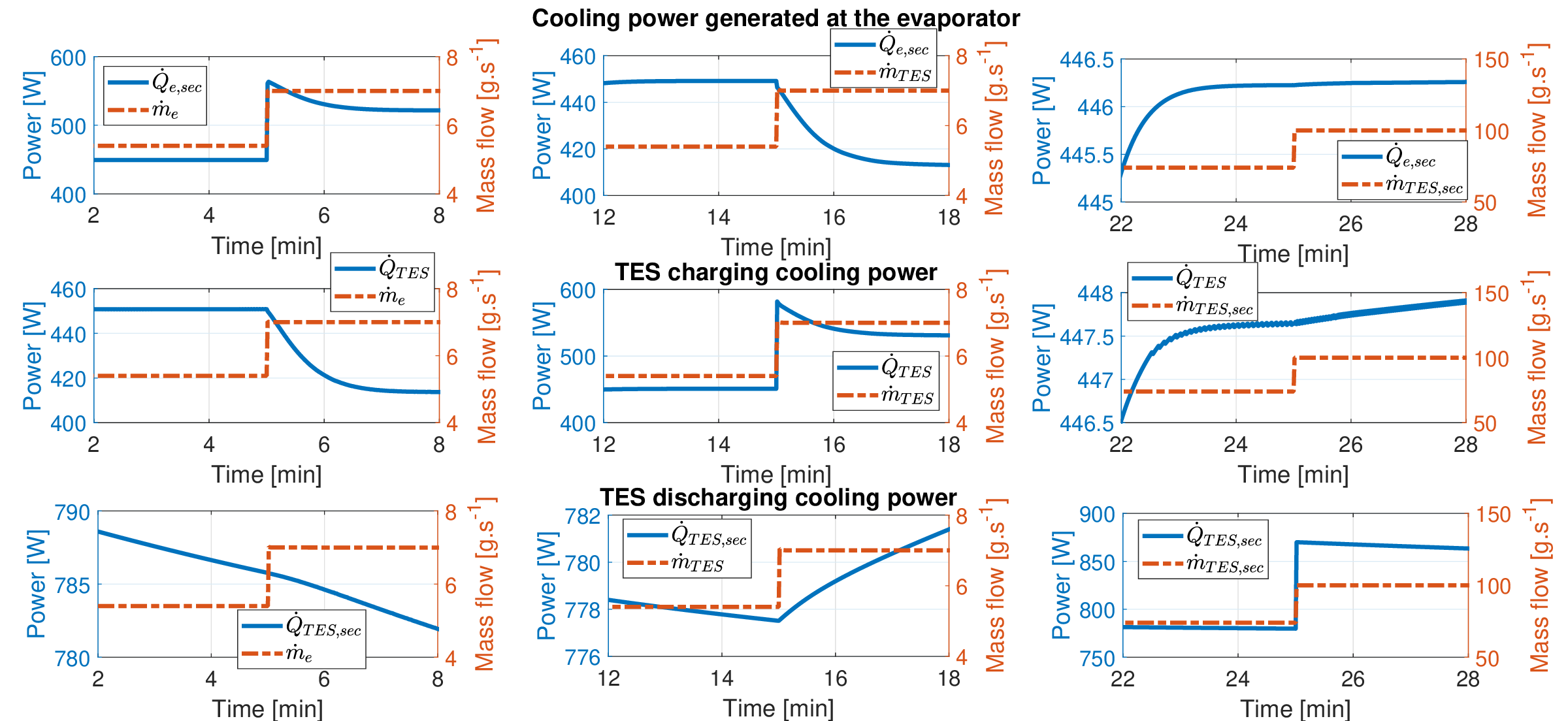}
	\caption{Multivariable step response of the system in operating mode 7 (cooling powers in absolute value).}
	\label{figStepResponse} 
\end{figure}

Once applied the supervisory control on $T_{SH}$ explained in subsection \ref{subSecOverview}, and considering the refrigerant mass flows $\dot{m}_e$ and $\dot{m}_{TES}$ as virtual manipulated variables, extensive simulations in open loop have been performed, in order to analyse the coupling between the cooling powers in every operating mode. Using the step-response method, the linear multivariable model shown in Equation Set \eqref{eq_Linear_Multivariable_Model} has been identified for the three main cooling powers ($\dot{Q}_{e,sec}$, $\dot{Q}_{TES}$, and $\dot{Q}_{TES,sec}$) with respect to the three mass flows involved ($\dot{m}_e$, $\dot{m}_{TES}$, and $\dot{m}_{TES,sec}$), for a given operating point in the most complex mode (\emph{Chamber + Charge + Discharge}, mode 7). The coupling analysis is focused on this mode since all relevant cooling powers are involved and the observed coupling appear also partially in the remaining operating modes. An example of the multivariable step response of the system is shown in Figure \ref{figStepResponse}, that is intended to serve as a qualitative representation of the most relevant coupling between the controlled variables.

\begin{equation}
	\begin{aligned}
		&\left[
		\begin{array}{c}
			\dot{Q}_{e,sec} (s) \\
			\dot{Q}_{TES} (s) \\
			\dot{Q}_{TES,sec} (s) \\
		\end{array}
		\right] = \bm{H}(s) 
		\left[
		\begin{array}{c}
			\dot{m}_e (s) \\
			\dot{m}_{TES} (s) \\
			\dot{m}_{TES,sec} (s) \\
		\end{array}
		\right]
		\\
		\bm{H}(s) &\approx 
		\left[
		\begin{array}{ccc}
			\frac{K_{11} (\tau_z s + 1)}{\tau_{dp} s + 1} & \frac{K_{12}}{\tau_{dp} s + 1} & K_{13} \\
			\frac{K_{21}}{\tau_{dp} s + 1} & \frac{K_{22}(\tau_z s + 1)}{\tau_{dp} s + 1} & K_{23} \\ 
			K_{31} & K_{32} & K_{33} \\
		\end{array}
		\right] = \\
		&= \left[
		\begin{array}{ccc}
			\frac{4.5 \cdot 10^4 (68 s + 1)}{42 s + 1} & \frac{-2 \cdot 10^4}{42 s + 1} & 0 \\
			\frac{-2 \cdot 10^4}{42 s + 1} & \frac{5 \cdot 10^4(68 s + 1)}{42 s + 1} & 0 \\ 
			0 & 0 & 0.3 \cdot 10^4 \\
		\end{array}
		\right]
	\end{aligned}
	\label{eq_Linear_Multivariable_Model}
\end{equation}

It is important to remark that, according to the cooling power sign criteria adopted in the previous modelling works by the authors \cite{Bejarano2017NovelSchemePCM,bejarano2018efficient}, the absolute values of all cooling powers have been represented in Figure \ref{figStepResponse}. It can be observed that, before any change in the refrigerant or secondary mass flows, the absolute value of the charging power $\dot{Q}_{TES}$ is lower than the absolute value of the discharging power $\dot{Q}_{TES,sec}$. It involves that the temperature of the intermediate fluid $T_{TES,int}$ increases slightly along the simulation, due to the energy balance in the intermediate fluid, even if no changes in the refrigerant or secondary mass flows are applied. Therefore, the discharging cooling power $\dot{Q}_{TES,sec}$ decreases slowly as $T_{TES,int}$ increases, as shown at the beginning of the left lowest plot, while the charging cooling power $\dot{Q}_{TES}$ increases slowly as $T_{TES,int}$ increases, as shown in the right middle plot. Moreover, in the latter plot, the increase in $\dot{Q}_{TES}$ is accelerated from $t$ = 25 min, due to the step change in $\dot{m}_{TES,sec}$, that involves that $T_{TES,int}$ increases a bit faster. Concerning the changes in the refrigerant and secondary flows, it is observed that the most important coupling takes place between $\dot{Q}_{e,sec}$ and $\dot{Q}_{TES}$, whereas the hypothesis of low coupling between the refrigerant circuit and the secondary fluid which circulates through the TES tank is confirmed and thus the corresponding elements in the transfer matrix $\bm{H}(s)$ have been neglected. 

Regarding the cooling power limits, the ranges of the actual and virtual manipulated inputs are indicated in Table \ref{tabInputRanges}. Given those ranges and applying the supervisory strategy on $T_{SH}$ described in subsection \ref{subSecOverview}, the cooling power ranges indicated in Table \ref{tabPowerConstraints} have been computed for every operating mode. It is important to note that the charging/discharging cooling power limits depend on the TES \emph{charge ratio} $\gamma_{TES}$, due to the increasing thermal resistance caused by the cylindrical shell in sensible zone, both in charging and discharging processes, that reduces the cooling power achievable for a given temperature difference between the cold/warm HTF and the PCM cylinder core, provided that it remains in latent zone \cite{Bejarano2017NovelSchemePCM}. Therefore, different values are provided in Table \ref{tabPowerConstraints}, considering that the freezing/melting boundary is located approximately on the edge, halfway, and at the centre of the PCM cylinder, which implies no thermal resistance due to sensible shell of the PCM cylinder, half thermal resistance, and the maximum thermal resistance, respectively.

\begin{table}[htbp]
	\centering
	\caption{Manipulated input ranges}
	\label{tabInputRanges}
	\scalebox{0.8}[0.8]{ \tabulinesep=0.5mm
		\begin{tabu} { L{7.5cm} C{2.2cm} C{2.7cm} C{1.2cm}}
			\toprule
			\emph{\textbf{Variable}} & \emph{\textbf{Symbol}} & \emph{\textbf{Range}} & \emph{\textbf{Unit}} \\ \midrule
			Expansion valve opening & $A_v$ & [10, 90] & \% \\
			TES expansion valve opening & $A_{v,TES}$ & [10, 90] & \% \\
			TES secondary mass flow	& $\dot{m}_{TES,sec}$ & [50, 440] & g s\textsuperscript{-1} \\
			Compressor speed & $N$ & [30, 50] & Hz \\
			\midrule
			Refrigerant mass flow at the evaporator & $\dot{m}_{e}$ & $\approx$ [1, 10] &g s\textsuperscript{-1} \\
			Refrigerant mass flow at the TES & $\dot{m}_{TES}$ & $\approx$ [1, 10] &g s\textsuperscript{-1}
			\\
			\bottomrule
			
	\end{tabu}}
\end{table}

\begin{table}[H]
	\centering
	\caption{Cooling power ranges}
	\label{tabPowerConstraints}
	\scalebox{0.8}[0.8]{ \tabulinesep=0.5mm
		\begin{tabu} {C{2.2cm} C{4.0cm} C{2.2cm} C{2.2cm} C{2.4cm}}
			\toprule

		\textbf{Operating mode}&\emph{Freezing/melting boundary location} & $\dot{Q}_{e,sec}$ [W] & $\dot{Q}_{TES}$ [W] & $\dot{Q}_{TES,sec}$ [W] \\ \toprule
					&Edge 	& [113, 759] & [103, 784] & -- \\
		\textbf{1}	&Halfway & [113, 759] & [103, 783] 	& -- \\
					&Centre	& [113, 759] & [103, 782] 	& -- \\
			\toprule
					&Edge 	& [139, 826] & -- & -- \\
		\textbf{2}	&Halfway & [139, 826] & -- & -- \\
					&Centre	& [139, 826] & -- & -- \\
			\toprule
					&Edge 	& [139, 826] & -- & [599, 1182] \\
		\textbf{3}	&Halfway & [139, 826] & -- & [458, 840] \\
					&Centre	& [139, 826] & -- & [300, 518] \\
			\toprule
					&Edge 	& -- & -- & [599, 1182] \\
		\textbf{4}	&Halfway & -- & -- & [458, 840] \\
					&Centre	& -- & -- & [300, 518] \\
			\toprule
					&Edge 	& -- & [135, 592] & -- \\
		\textbf{5}	&Halfway & -- & [135, 589] 	& -- \\
					&Centre	& -- & [135, 587] 	& -- \\
			\toprule
					&Edge 	& -- & [139, 613] & [646, 1262] \\
		\textbf{6}	&Halfway & -- & [139, 615] & [608, 1185] \\
					&Centre	& -- & [139, 617] & [572, 1114] \\
			\toprule
					&Edge 	& [108, 771] & [101, 816] & [648, 1308] \\
		\textbf{7}	&Halfway & [108, 771] & [102, 813] & [608, 1240] \\
					&Centre	& [108, 771] & [102, 814] & [577, 1178] \\
			\bottomrule
	\end{tabu}}
\end{table} 

\begin{figure}[h]
	\centering
	\includegraphics[width=7cm]{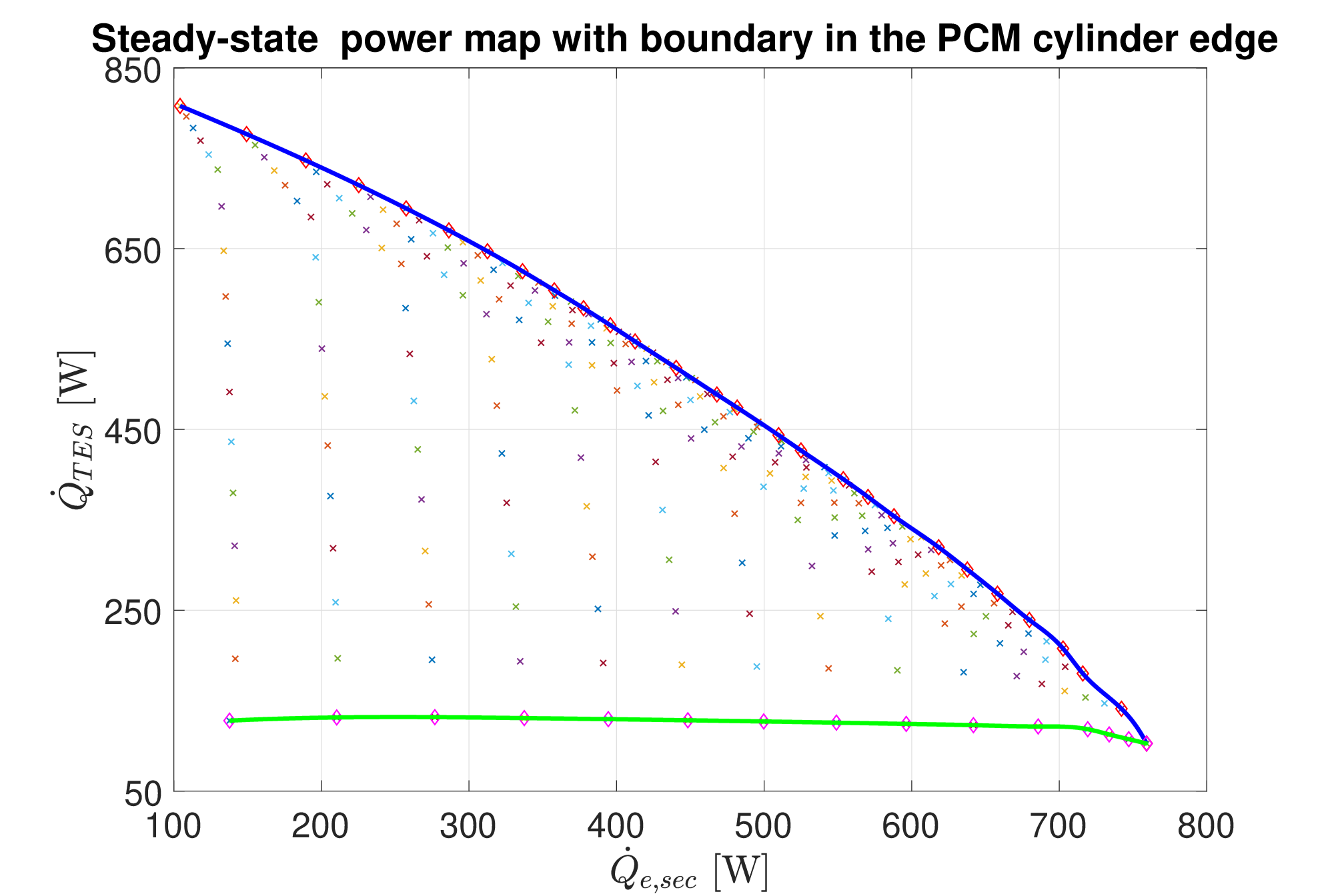}
	\caption{Steady-state cooling power map for operating mode 1 when the freezing boundary locates on the PCM cylinder edge.} 
	\label{figCoolingPowersMode1_edge}
\end{figure}

It is important to note that, in operating modes 1 and 7, $\dot{Q}_{e,sec}$ and $\dot{Q}_{TES}$ are highly coupled, since two different refrigerant flows merge at the compressor intake (point A in Figure \ref{figEsquemaCicloPCM}), coming from the evaporator and the TES tank. Thus, the ranges detailed in Table \ref{tabPowerConstraints} for those modes might not be realistic, since they include some unachievable points. For instance, regarding mode 1, Figure \ref{figCoolingPowersMode1_edge} shows the steady-state cooling power map when the freezing boundary locates in the PCM cylinder edge.  

In Figure \ref{figCoolingPowersMode1_edge} the continuous lines stand for the overall limits, while the small crosses represent admissible points in steady state. Similar qualitative results can be obtained for different locations of the freezing boundary. Therefore, it can be concluded that, concerning the maximum power limits, the greater $\dot{Q}_{e,sec}$, the smaller $\dot{Q}_{TES}$. It involves that for a given cooling demand which might need to be satisfied only at the evaporator (mode 1), the maximum achievable charging power is constrained by a curve similar to that represented in Figure \ref{figCoolingPowersMode1_edge}.  

\subsection{Controller design} \label{subSecCoolingPowerControl}

\begin{figure} [h]
	\begin{center}
		\includegraphics[width=13cm,trim = 70 535 0 120,clip]{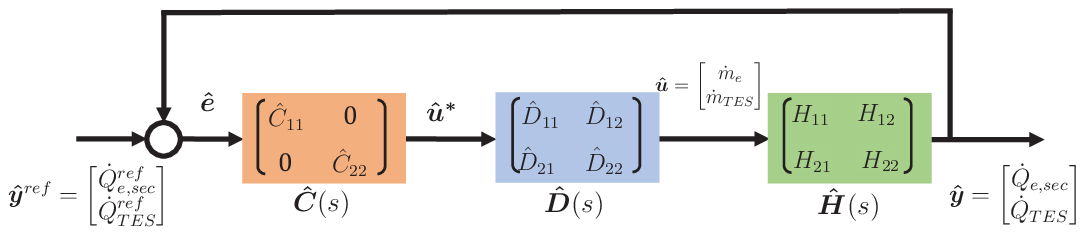}   
		\caption{Partial decoupling strategy.} 
		\label{figDesacoplo}
	\end{center}
\end{figure}

Given the high coupling between $\dot{Q}_{e,sec}$ and $\dot{Q}_{TES}$, a decoupling strategy is proposed for controlling both cooling powers, leaving $T_{SH}$ and $\dot{Q}_{TES,sec}$ out of the decoupling network. A partial decoupling matrix $\bm{\hat{D}}(s)$ is inserted between the partial controller matrix $\bm{\hat{C}}(s)$ and the partial system matrix $\bm{\hat{H}}(s)$, as shown in Figure \ref{figDesacoplo}. Every transfer function $\hat{D}_{ij}(s) \;\; \forall i,j = 1,2$ is intended to be conveniently computed so that the design of the desired diagonal controller matrix $\bm{\hat{C}}(s)$ with transfer functions $\hat{C}_{11}(s)$ and $\hat{C}_{22}(s)$, can be addressed with minimum crossed interaction.

The input-output matching shown in Table \ref{tabControlPairing} is selected, according to the analysis performed using the \emph{Relative Gain Array} (RGA) method \cite{Bristol1966}, where the relative gain matrix $\bm{\lambda}$ is calculated from the system static gains $K_{ij}$ and the closed-loop ones $K_{C_{ij}} \; \forall i,j = 1,2$, as indicated in Equation Set \eqref{eq_RGA}. It is shown that the selected input-output matching achieves the elements closest to one in the relative gain matrix $\bm{\lambda}$.

\begin{table} [htbp]
	\centering
	\caption{Input-output matching applied in the decentralised cooling power controller}
	\label{tabControlPairing}
	\scalebox{0.8}[0.8]{ \tabulinesep=0.5mm
		\begin{tabu} { C{3.9cm} C{5.5cm} C{3.8cm} }
			\toprule
			\emph{\bf Manipulated input} & \emph{\bf Virtual manipulated input} & \emph{\bf Controlled output} \\ 
			\midrule
			$A_v$ & $\dot{m}_e$ & $\dot{Q}_{e,sec}$ \\ 
			\midrule
			$A_{v,TES}$ & $\dot{m}_{TES}$ & $\dot{Q}_{TES}$ \\ 
			\midrule
			$\dot{m}_{TES,sec}$ & -- & $\dot{Q}_{TES,sec}$ \\ 
			\midrule
			$N$ & -- & $T_{SH}$ \\ 
			\bottomrule
	\end{tabu}}
\end{table}

\begin{equation} 
	\begin{aligned}
	\bm{\hat{K}} = 
	\left[
	\begin{matrix}
		{K_{11}} & {K_{12}} \\
		{K_{21}} & {K_{22}} \\
	\end{matrix}
	\right] &=
	\left[
	\begin{matrix}
		4.5 & -2.0\\
		-2.0 & 5.0\\
	\end{matrix}
	\right] \cdot 10^{4} \\
	\left[
	\begin{matrix}
		\frac{1}{K_{C_{11}}} & \frac{1}{K_{C_{12}}}\\
		\frac{1}{K_{C_{21}}} & \frac{1}{K_{C_{22}}}\\
	\end{matrix}
	\right] &=
	\left(
	\left[
	\begin{matrix}
		K_{11} & K_{12} \\
		K_{21} & K_{22} \\
	\end{matrix}
	\right]^{-1}
	\right)^T \\
	\bm{\lambda} =
	\left[
	\begin{matrix}
		\frac{K_{11}}{K_{C_{11}}} & \frac{K_{12}}{K_{C_{12}}} \\
		\frac{K_{21}}{K_{C_{21}}} & \frac{K_{22}}{K_{C_{22}}} \\
	\end{matrix}
	\right] &= 
	\left[
	\begin{matrix}
		1.22 & -0.22 \\
		-0.22 & 1.22\\
	\end{matrix}
	\right]
	\end{aligned}
	\label{eq_RGA}
\end{equation}

Once defined the input-output matching, the decoupling matrix $\bm{\hat{D}}(s)$ is calculated in such a way that satisfies Equation \eqref{eq_decoupling}, where $\bm{\hat{K}}^{diag}$ is the desired diagonal matrix to be used within the design of the monovariable controllers $\hat{C}_{11}(s)$ and $\hat{C}_{22}(s)$. It is important to note that, given the fast cooling power dynamics in comparison with the time scale of the scheduling strategy, the decoupling matrix and the decentralised controller are intended to be calculated from the static gain matrix $\bm{\hat{K}}$, rather than the low-order linear transfer matrix $\bm{\hat{H}}(s)$.

\begin{equation} 
	\bm{\hat{K}} \cdot \bm{\hat{D}} = 
	\left[
	\begin{matrix}
		K_{11} & K_{12}\\
		K_{21} & K_{22}\\
	\end{matrix}
	\right]
	\left[
	\begin{matrix}
		\hat{D}_{11} & \hat{D}_{12}\\
		\hat{D}_{21} & \hat{D}_{22}\\
	\end{matrix}
	\right] = 
	\bm{\hat{K}}^{diag} (s) =
	\left[
	\begin{matrix}
		\hat{K}_{11}^{diag} & 0 \\
		0 & \hat{K}_{22}^{diag}\\
	\end{matrix} 
	\right]
	\label{eq_decoupling}
\end{equation}

A suitable solution for the decoupling matrix $\bm{\hat{D}}$ is provided in Equation Set \eqref{eq_Dsol}, where the resulting diagonal matrix $\bm{\hat{K}}^{diag}$ is also shown.

\begin{equation} 
	\begin{aligned}
		\bm{\hat{D}} &= \dfrac{adj(\bm{\hat{K}})}{det(\bm{\hat{K}})} \cdot \bm{\hat{K}}^{diag}\\
		\bm{\hat{D}} &= 
		\left[ 
		\begin{matrix}
			1 & \frac{adj(K_{12})}{adj(K_{22})} \\
			\frac{adj(K_{21})}{adj(K_{11})} & 1 \\
		\end{matrix}
		\right] = 
		\left[
		\begin{matrix}
			1 & 0.44 \\
			0.4 & 1\\
		\end{matrix} 
		\right] \\
		\bm{\hat{K}}^{diag} &= 
		\left[ 
		\begin{matrix}
			\frac{det(\bm{\hat{K}})}{adj(K_{11})} & 0\\
			0 & \frac{det(\bm{\hat{K}})}{adj(K_{22})}\\
		\end{matrix}
		\right] = 
		\left[
		\begin{matrix}
			3.7 & 0 \\
			0 & 4.1 \\
		\end{matrix} 
		\right] \cdot 10^{4} \\
	\end{aligned}
	\label{eq_Dsol}
\end{equation}

\begin{figure} [h]
	\begin{center}
		\includegraphics[width=13cm,trim = 70 535 0 120,clip]{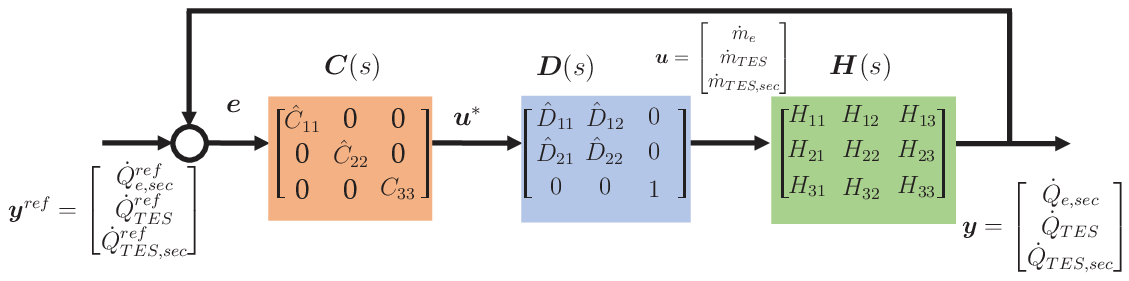}   
		\caption{Cooling power control strategy.} 
		\label{figDesacoplo_3}
	\end{center}
\end{figure}

Eventually, once calculated the partial decoupling matrix $\bm{\hat{D}}$, the partial decentralised cooling power controller is computed according to the diagonal static gain matrix $\bm{\hat{K}}^{diag}$. Two PI regulators are applied to both independent cooling power control loops, whereas a logical network is also added to override and/or reset the controllers when needed, according to expected operating mode changes due to the cooling power references. All controller parameters are detailed in Table \ref{tabControlParameters}, where the low-level controller parameters regarding the refrigerant mass flows $\dot{m}_e$ and $\dot{m}_{TES}$ and the degree of superheating $T_{SH}$ have been also included. Note that the complete controller transfer matrix $\bm{C}(s)$ is applied as shown in Figure \ref{figDesacoplo_3}, where the supervisory strategy on the degree of superheating is also applied by manipulating the compressor speed $N$ when necessary.

\begin{table}[h]
	\centering
	\caption{Controller parameters} 
	\label{tabControlParameters}
	\scalebox{0.9}[0.9]{ \tabulinesep=0.5mm
	\begin{tabular}{c c c}
		\toprule
		\multirow{2}{*}{\textbf{Controller}} & \multicolumn{2}{c}{\textbf{PI parameters}} \\
		& $K_p$ & $T_i$ \\
		\toprule
		$C_{11}(s)$ & $\frac{10^{-4}}{23}$ kg s\textsuperscript{-1} W\textsuperscript{-1} & 1.5 s \\
		\bottomrule
		$C_{22}(s)$ & $\frac{10^{-4}}{23}$ kg s\textsuperscript{-1} W\textsuperscript{-1} & 1.5 s \\
		\bottomrule
		$C_{33}(s)$ & 10\textsuperscript{-4} kg s\textsuperscript{-1} W\textsuperscript{-1} & 2 s \\
		\toprule
		$C_{\dot{m}_e}(s)$ & $10^{3}$ \% kg\textsuperscript{-1} s & 1 s \\
		\bottomrule
		$C_{\dot{m}_{TES}}(s)$ & $10^{3}$ \% kg\textsuperscript{-1} s & 1 s \\
		\bottomrule
		$C_{T_{SH}} (s)$ & 1.33 Hz K\textsuperscript{-1} & 1.05 s \\
		\bottomrule
	\end{tabular}}	
\end{table}

\subsection{Simulation results} \label{subSecSimulationResults}

\begin{figure}[H]
	\centering
	\subfigure[Control on the cooling power generated at the evaporator.]{
		\includegraphics[width=7.0cm] {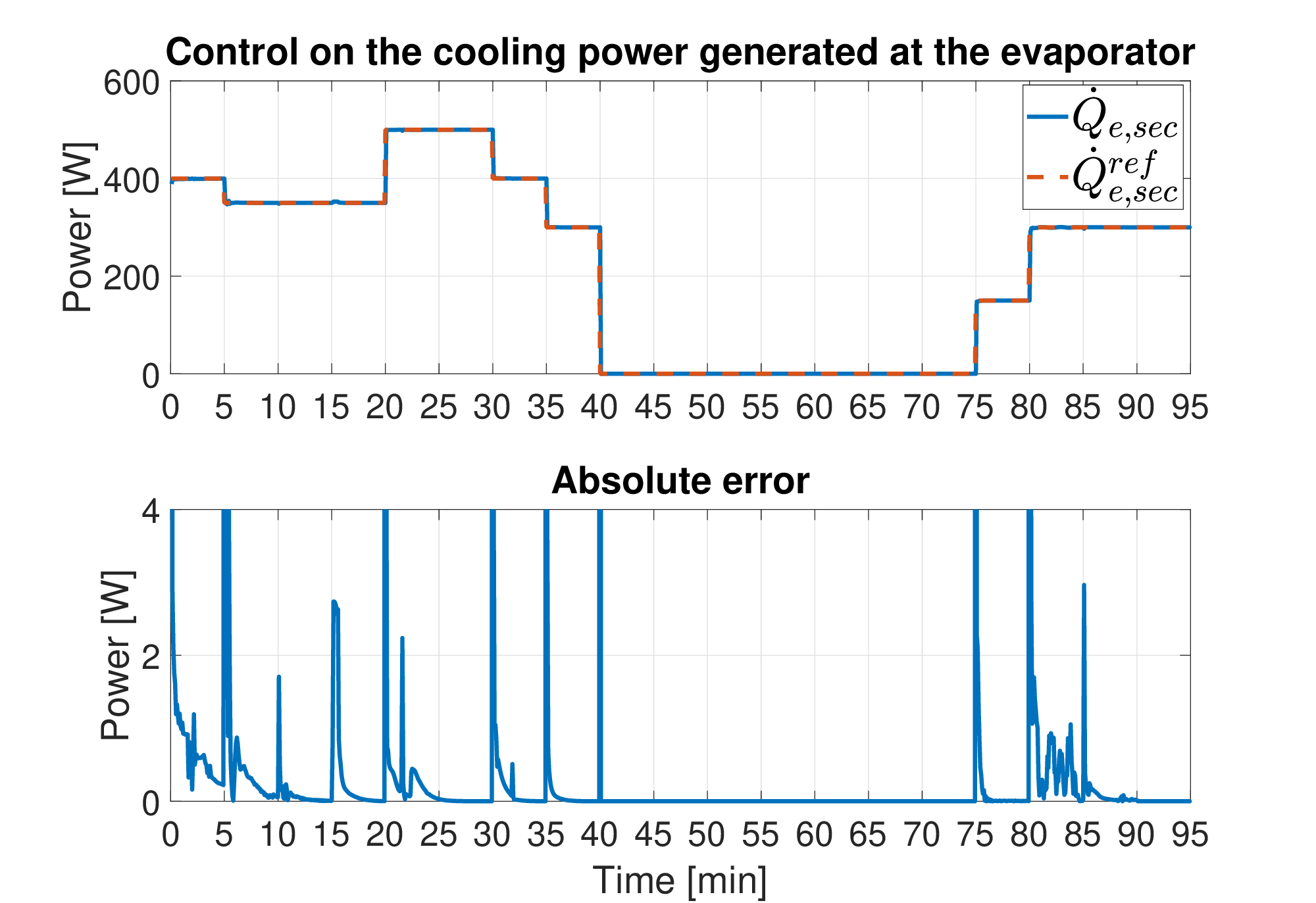}
		\label{figCoolingPowerControl_Q_e_sec}
	}\subfigure[Control on the cooling power transferred to the TES tank.]{
		\includegraphics[width=7.0cm] {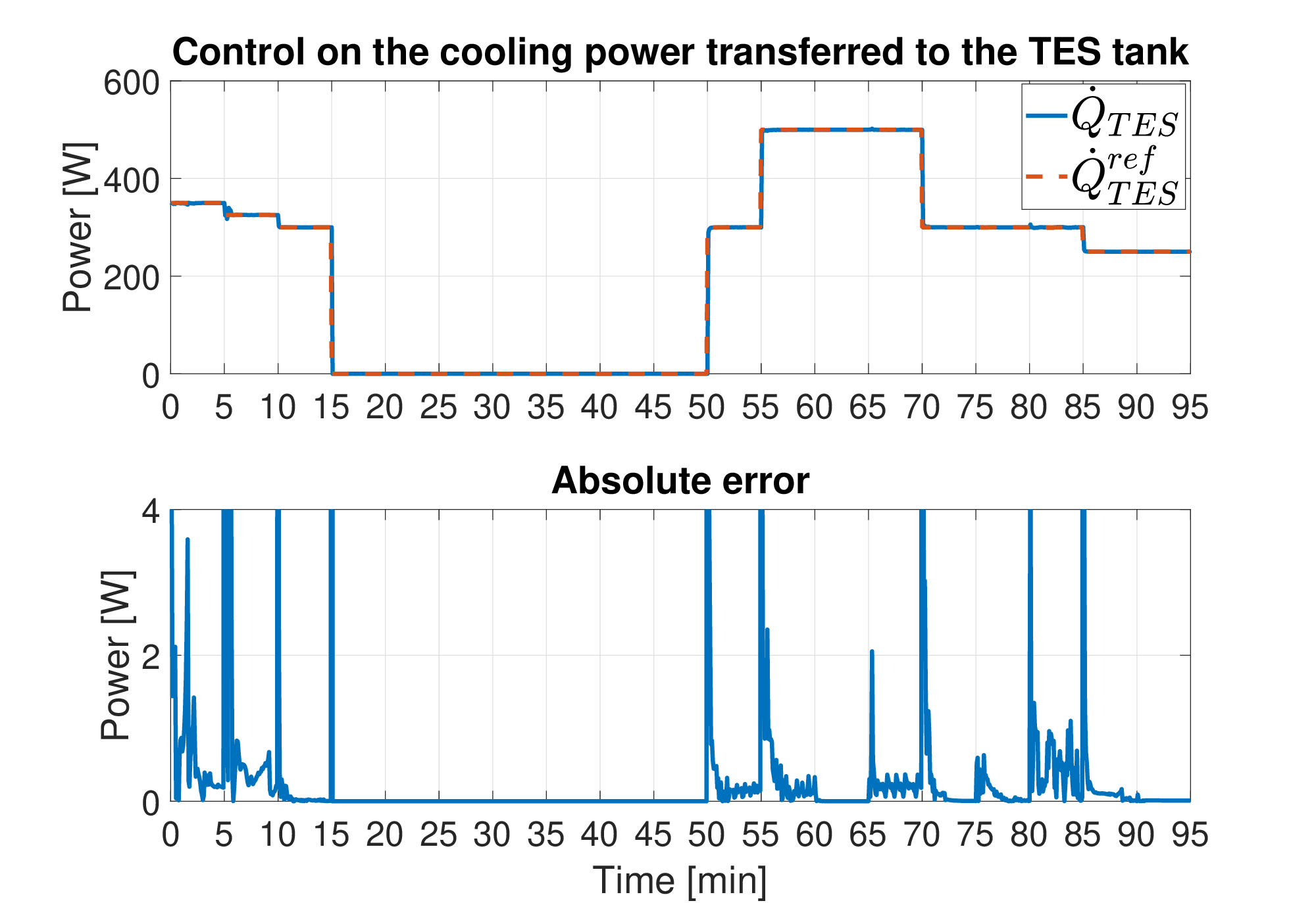}
		\label{figCoolingPowerControl_Q_TES}
	}
	\subfigure[Control on the cooling power released from the TES tank.]{
		\includegraphics[width=7.0cm] {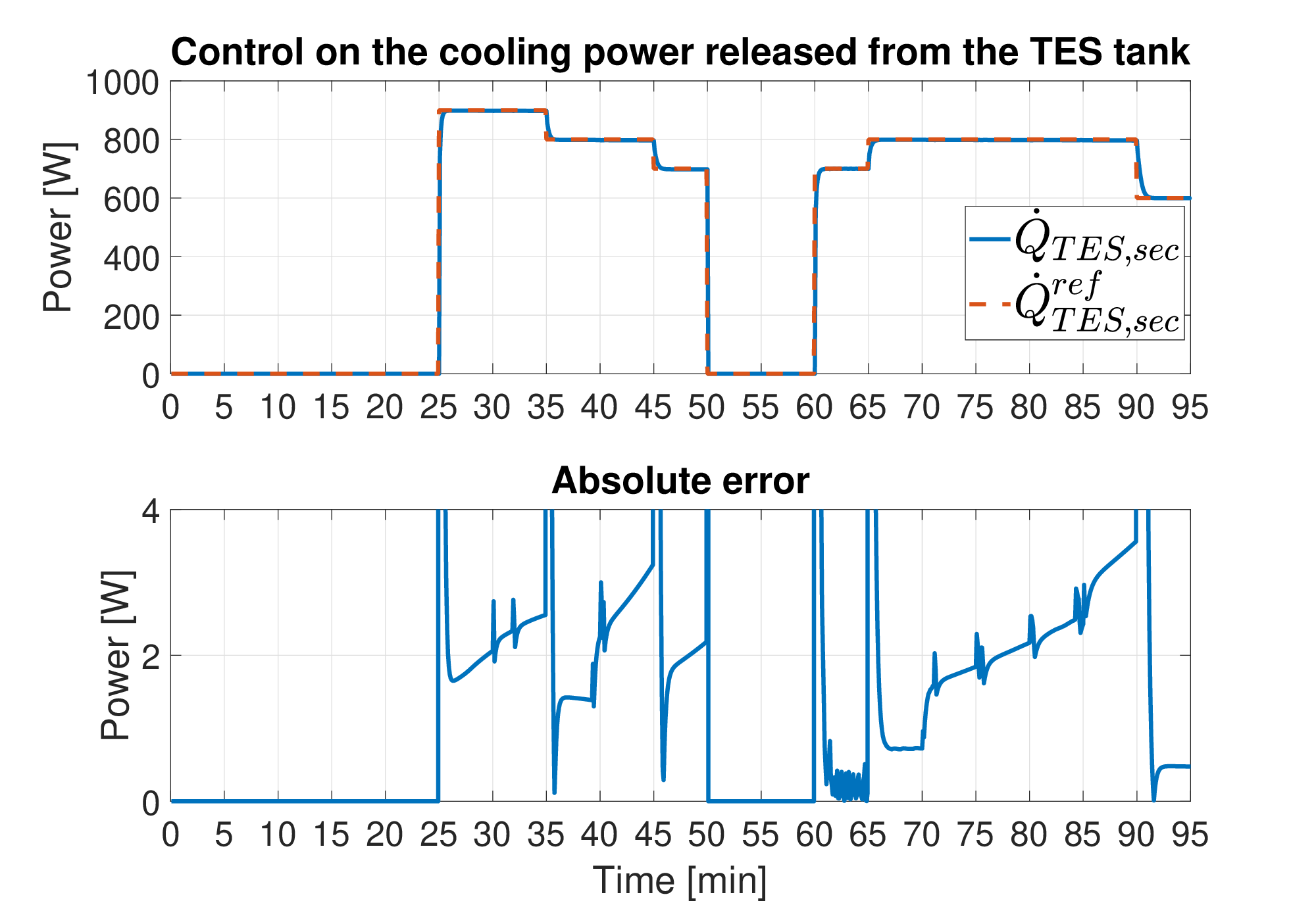}
		\label{figCoolingPowerControl_Q_TES_sec}
	}\subfigure[Operating mode profile.]{
		\includegraphics[width=7.0cm] {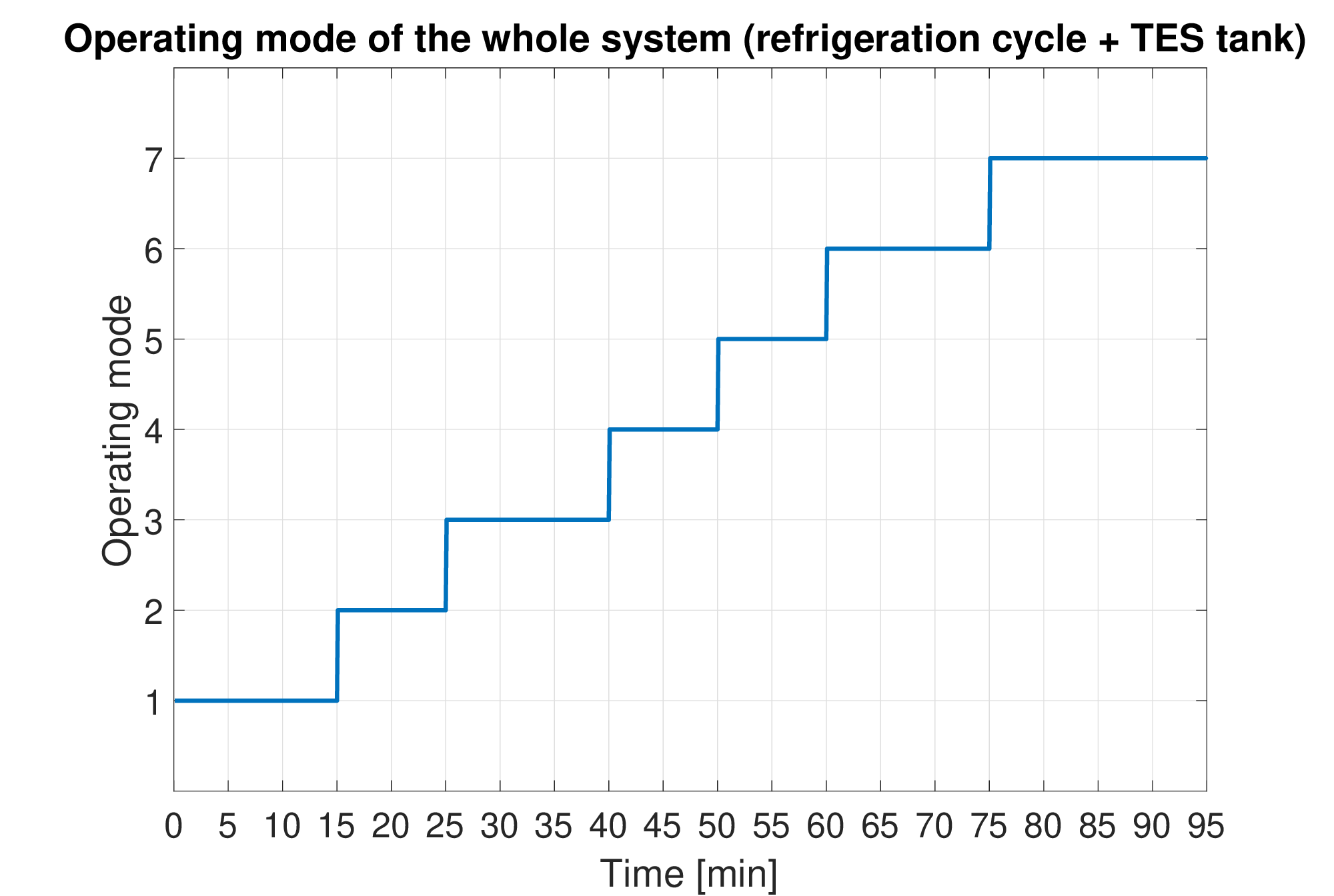}
		\label{figCoolingPowerControl_op_mode}
	}
	\caption{Cooling power controller performance and operating mode profile.}
	\label{figCoolingPowerControl_Q_OP}				
\end{figure}

\begin{figure}[H]
	\centering
	\subfigure[Degree of superheating at the compressor intake.]{
		\includegraphics[width=7.0cm] {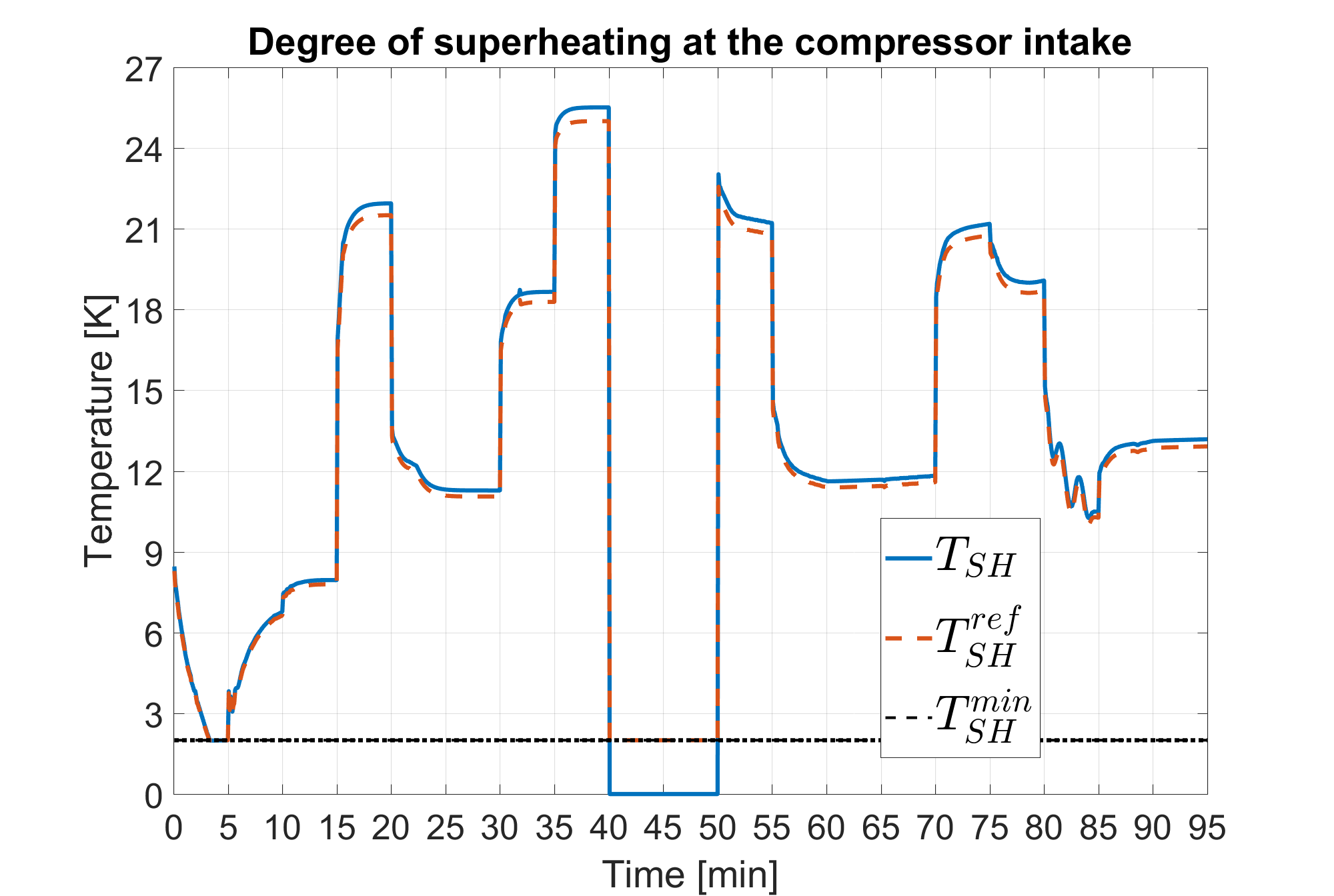}
		\label{figCoolingPowerControl_TSH}
	}\subfigure[Compressor speed.]{	
		\includegraphics[width=7.0cm] {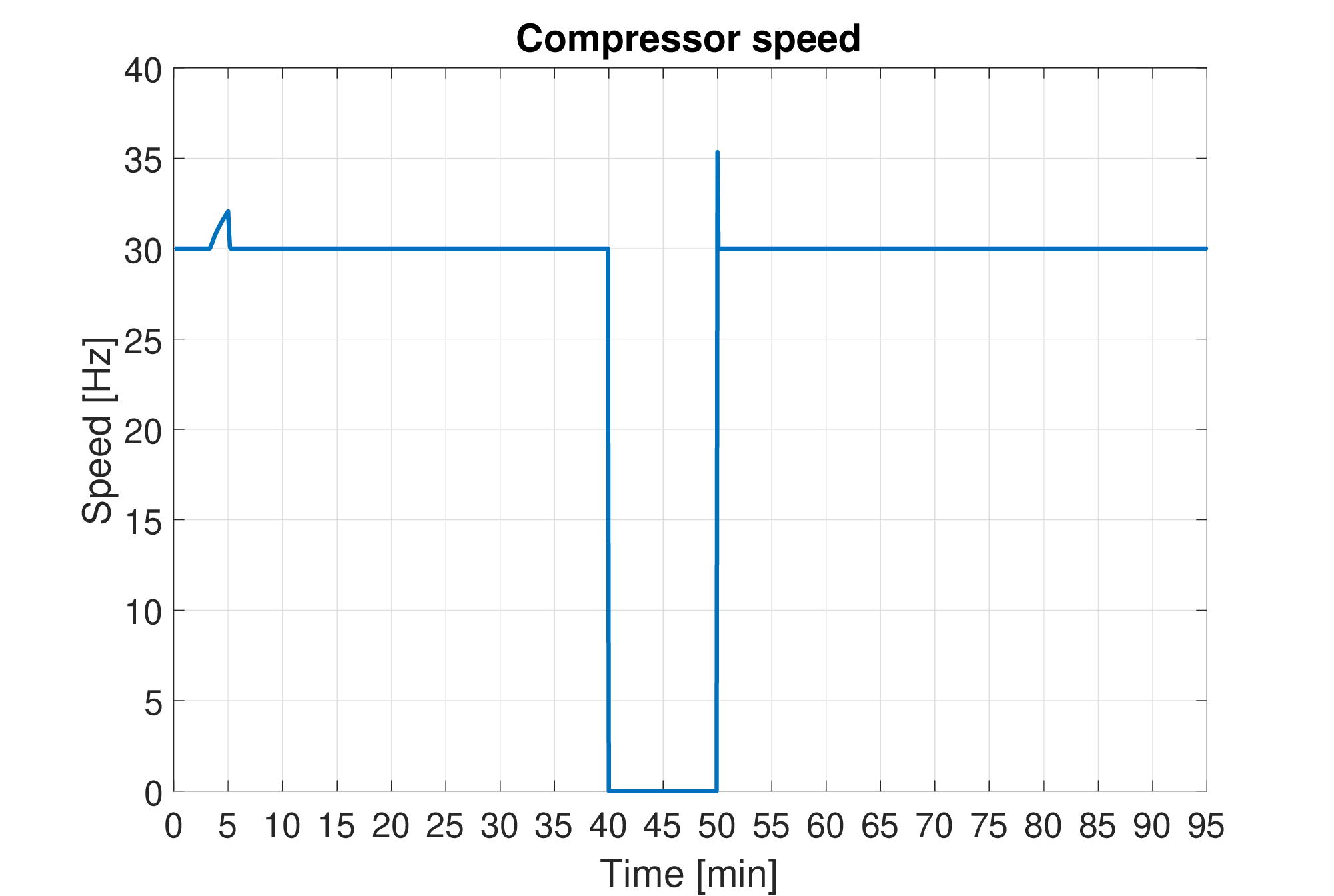}
		\label{figCoolingPowerControl_N}
	}
	\caption{Supervisory control on $T_{SH}$.}
	\label{figCoolingPowerControl_TSH_N}	
\end{figure}

\begin{figure}[H]
	\centering
	\subfigure[Refrigerant mass flow distribution.]{
		\includegraphics[width=7.0cm] {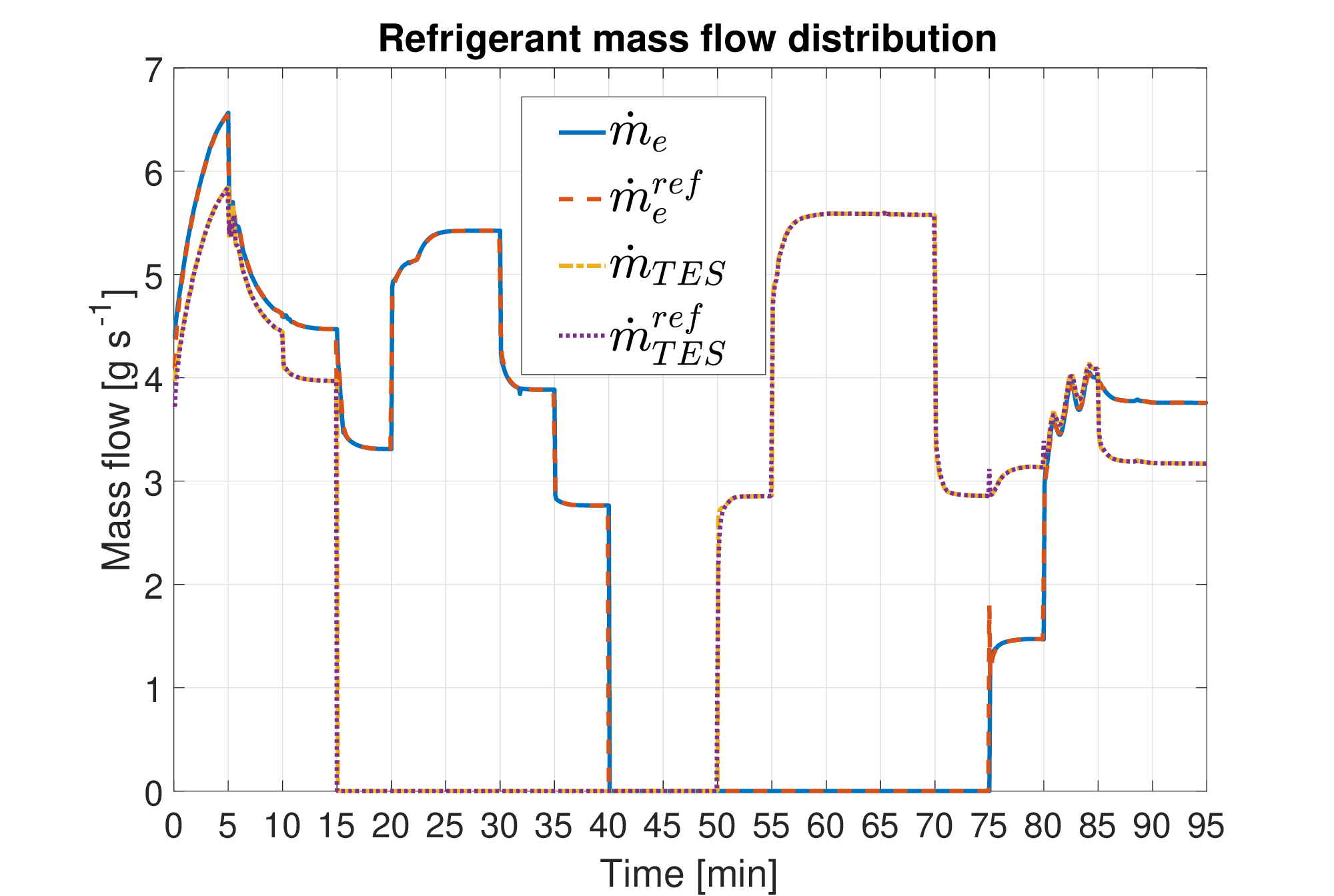}
		\label{figCoolingPowerControl_mR}
	}\subfigure[Expansion valve opening.]{
		\includegraphics[width=7.0cm] {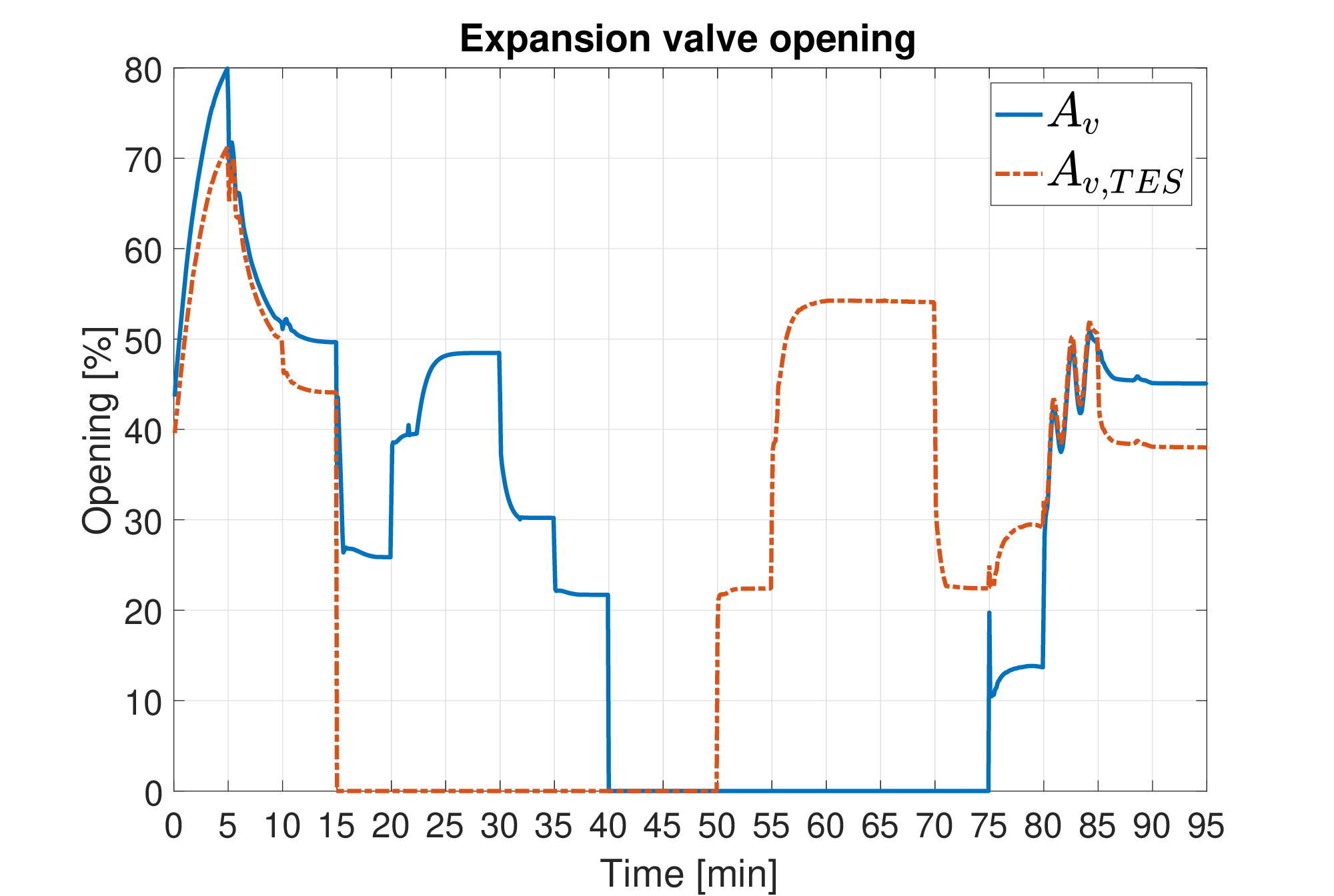}
		\label{figCoolingPowerControl_Av}
	}
	\caption{Refrigeration mass flow control.}
	\label{figCoolingPowerControl_mR_Av}	
\end{figure}

\begin{figure}[H]
	\centering
	\subfigure[Secondary mass flow circulating through the TES tank.]{
		\includegraphics[width=7.0cm] {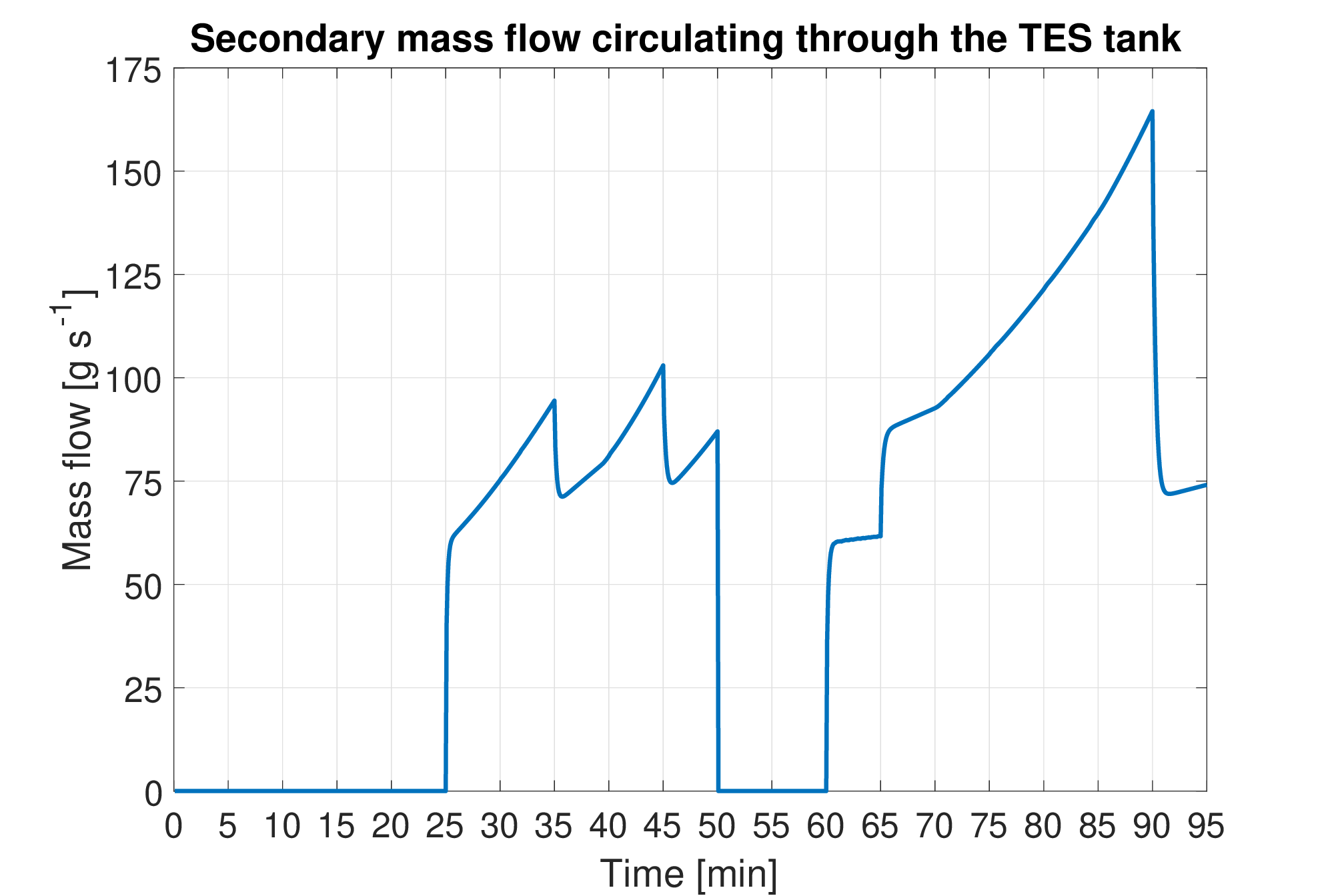}
		\label{figCoolingPowerControl_m_TES_sec}
	}\subfigure[TES tank \emph{charge ratio}.]{
		\includegraphics[width=7.0cm] {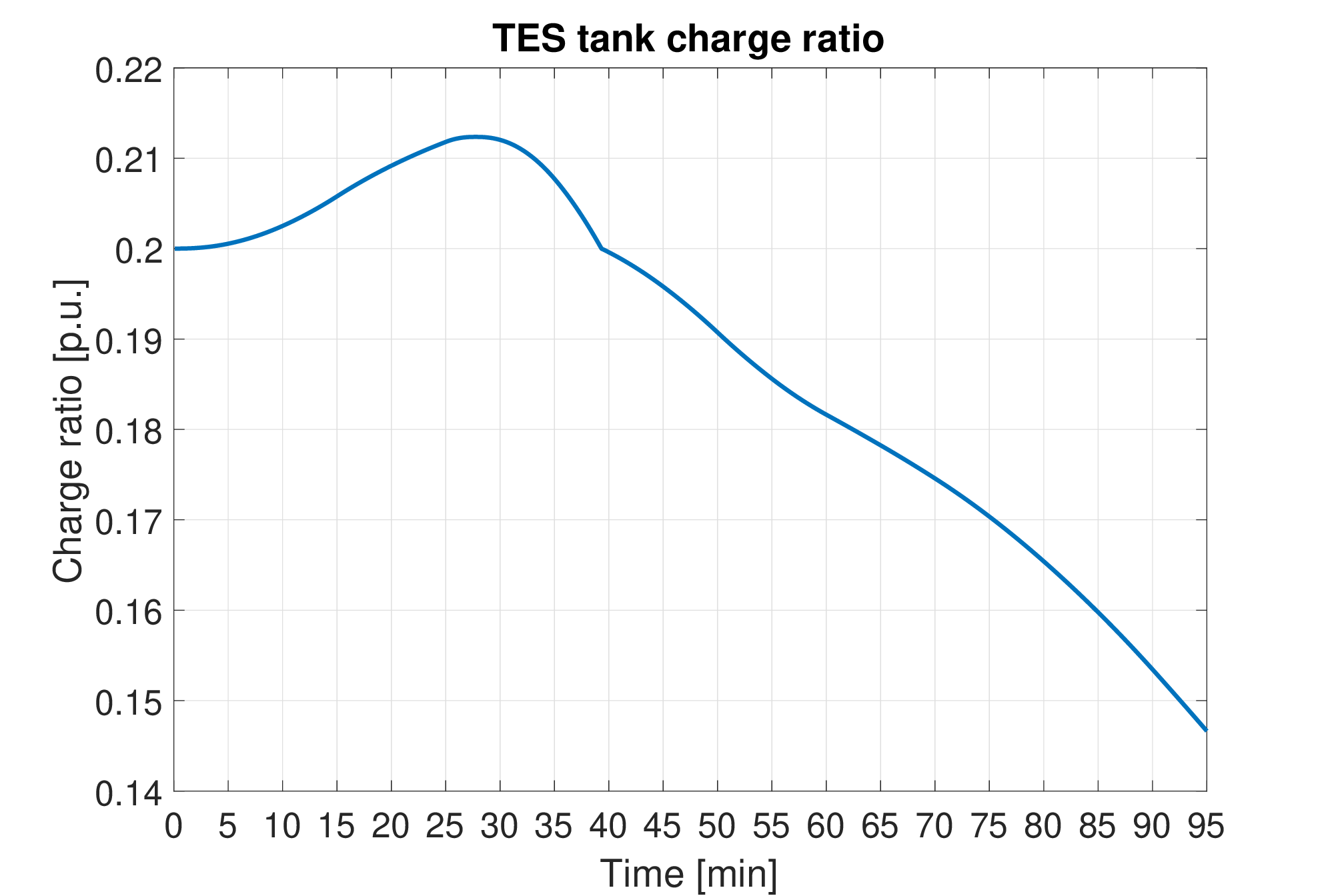}
		\label{figCoolingPowerControl_gamma_TES}
	}
	\caption{Secondary mass flow circulating through the TES tank and \emph{charge ratio}.} 
	\label{figCoolingPowerControl_other}
\end{figure}

Figures \ref{figCoolingPowerControl_Q_OP} -- \ref{figCoolingPowerControl_other} show some simulation results of the cooling power control when exploring diverse operating modes. The references on the cooling powers are imposed considering the limits explained in subsection \ref{subSecCoupling} and indicated in Table \ref{tabPowerConstraints}, in such a way that the cooling power control can be tested in all operating modes. The sampling time is 5 s, which represents the baseline sampling rate of the whole control strategy.  

Then, Figure \ref{figCoolingPowerControl_Q_OP} represents the control performance for $\dot{Q}_{e,sec}$, $\dot{Q}_{TES}$, and $\dot{Q}_{TES,sec}$, along with the operating mode profile resulting from the imposed power references. Moreover, the performance of the supervisory control on the degree of superheating at the compressor intake is represented in Figure \ref{figCoolingPowerControl_TSH_N}, along with the corresponding manipulated variable $N$. Figure \ref{figCoolingPowerControl_mR_Av} shows the refrigerant mass flow distribution between the evaporator and the TES tank, when applicable, as well as the actual opening of both expansion valves. Eventually, Figure \ref{figCoolingPowerControl_other} shows the secondary mass fluid $\dot{m}_{TES,sec}$ and the \emph{charge ratio} along the simulation.

It is observed in Figure \ref{figCoolingPowerControl_Q_OP} that alternative step changes in the set points of the corresponding non-zero cooling powers have been applied for all operating modes, and the decoupling strategy shows to effectively reduce the coupling between the cooling powers while the controller provides very fast reference tracking. Indeed, given the intended sampling time of the scheduling strategy (in the order of several minutes), the achieved reference tracking turns out to be fast enough to meet the speed requirements. Regarding the supervision of the degree of superheating, it is shown in Figure \ref{figCoolingPowerControl_TSH_N} that only before $t$ = 5 min the compressor speed $N$ must be increased to ensure that $T_{SH}$ does not drop below $T_{SH}^{min}$ while providing the required cooling demand at the evaporator and charging the TES tank with the required cooling power.

Concerning the TES tank \emph{charge ratio}, it evolves while charging and discharging powers are applied, as shown in Figure \ref{figCoolingPowerControl_gamma_TES}. During the first 15 min the system is operating in mode 1, thus the refrigerant circulates through the TES tank while charging it. However, even when the operating mode switches to mode 2, the TES tank continues to be charged due to the thermal inertia of the intermediate fluid, until some time after $t$ = 25 min, when the system switches to mode 3. At this moment the TES tank starts to be discharged, and the \emph{charge ratio} continues to decrease during the operation in mode 4. In mode 5, the TES tank is being charged, but due to the thermal inertia of the previous discharging process the temperature of the intermediate fluid is still over the PCM phase-change temperature $T_{pcm}^{lat}$ and it continues to be discharged during all the corresponding time interval. In mode 6, the TES tank is being charged and discharged at the same time, but $\dot{Q}_{TES}^{ref}$ is smaller than $\dot{Q}_{TES,sec}^{ref}$, as shown in Figures \ref{figCoolingPowerControl_Q_TES} and \ref{figCoolingPowerControl_Q_TES_sec}, thus the net effect on $\gamma_{TES}$ is negative and it continues to decrease. This net effect persists in mode 7 despite the step changes on $\dot{Q}_{TES}^{ref}$ and $\dot{Q}_{TES,sec}^{ref}$, thus $\gamma_{TES}$ keeps a decreasing trend until the end of the simulation.

Note also that the fact that the TES tank charges or discharges affects the related manipulated variables. For instance, regarding the discharging process, let analyse in depth the time interval $t \in$ [25,35] min. The operating mode imposed by the cooling power references is mode 3, where the secondary fluid is cooled both at the evaporator and at the TES tank, which is meanwhile discharged. The reference on $\dot{Q}_{TES,sec}$ remains constant, as represented in Figure \ref{figCoolingPowerControl_Q_TES_sec}, but Figure \ref{figCoolingPowerControl_m_TES_sec} shows that the corresponding control action $\dot{m}_{TES,sec}$ is gradually increased during the time interval under consideration. It is due to the increasing thermal resistance caused by the inward growing shell in sensible zone of the PCM cylinders, that forces the control action to increase in order to transfer the same cooling power $\dot{Q}_{TES,sec}$. Similar effects can also appear during the TES charging process, which leads all refrigerant variables to vary in order to keep the transferred charging cooling power $\dot{Q}_{TES}$ constant, but as analysed in subsection \ref{subSecTESTank}, this effect is much less significant in the case of the charging process.


\section{Conclusions and future work} \label{secConclusions}

In this article, a novel hybrid setup including a PCM-based TES tank that complements a vapour-compression refrigeration facility has been modelled and analysed. The TES tank is intended to be used as an energy buffer which allows the decoupling of cold-energy production and demand and it may help reduce operating cost. The combined plant has been described, paying special attention to the TES tank and its interconnection with the refrigeration cycle. 

The modelling of every element has been addressed separately, while a combined model has been developed, describing the different time scales arisen from a frequency analysis: faster dynamics are linked to the refrigeration cycle, while the slower ones are caused by the heat transfer within the TES tank. Moreover, eight operating modes have been defined according to all possible combinations of the activation state of the main cooling powers involved ($\dot{Q}_{TES}$, $\dot{Q}_{TES,sec}$, and $\dot{Q}_{e,sec}$), while the system statics in all operating modes have been calculated, giving rise to combined power maps that show the high coupling between the aforementioned cooling powers in most operating modes. 

A decentralised strategy has been proposed for the cooling power control, including a decoupling matrix, an heuristic algorithm for the control of the degree of superheating, and a cascade strategy for the expansion valve controllers. Eventually, a thorough simulation on the model previously described has been presented, where the selected references on the three main cooling powers cause all operating modes to be tested. The decoupling strategy shows to effectively reduce the coupling between the cooling powers, especially $\dot{Q}_{e,sec}$ and $\dot{Q}_{TES}$, while the controller provides very fast reference tracking, fulfilling the speed requirements given by the intended sampling time of the forthcoming scheduling strategy, in the order of several minutes. Moreover, the consequences of the variations of the TES tank \emph{charge ratio} along the simulation on the manipulated variables have been discussed and justified. 

In conclusion, in this work the cooling power loops have been studied in detail and the power tracking control has been solved, which enables to address the aimed scheduling strategy that calculates the references on the main cooling powers, where diverse issues are intended to be considered, i.e. real-time cooling demand satisfaction, economic, efficiency, and feasibility criteria. Moreover, as future work, the simulation results provided by the developed model are intended to be validated as soon as the experimental facility is completely operative.  


\section*{Acknowledgements}

The authors would like to acknowledge Spanish MCeI (Grants DPI2015-70973-R and DPI2016-79444-R) for funding this work, as well as University of Seville through VI PPIT-US program. The cooperation of INESC-ID was supported by FCT (Portugal) under UID/CEC/50021/2019.



\bibliographystyle{IEEEtran}
\bibliography{bibliography}

\begin{thebibliography}{10}
\providecommand{\url}[1]{#1}
\csname url@samestyle\endcsname
\providecommand{\newblock}{\relax}
\providecommand{\bibinfo}[2]{#2}
\providecommand{\BIBentrySTDinterwordspacing}{\spaceskip=0pt\relax}
\providecommand{\BIBentryALTinterwordstretchfactor}{4}
\providecommand{\BIBentryALTinterwordspacing}{\spaceskip=\fontdimen2\font plus
\BIBentryALTinterwordstretchfactor\fontdimen3\font minus
  \fontdimen4\font\relax}
\providecommand{\BIBforeignlanguage}[2]{{%
\expandafter\ifx\csname l@#1\endcsname\relax
\typeout{** WARNING: IEEEtran.bst: No hyphenation pattern has been}%
\typeout{** loaded for the language `#1'. Using the pattern for}%
\typeout{** the default language instead.}%
\else
\language=\csname l@#1\endcsname
\fi
#2}}
\providecommand{\BIBdecl}{\relax}
\BIBdecl

\bibitem{Rasmussen2005}
B.~P. Rasmussen, A.~Musser, and A.~G. Alleyne, ``Model-driven system
  identification of transcritical vapor compression systems,'' \emph{IEEE
  Trans. Control Syst. Technol.}, vol.~13, pp. 444--451, 2005.

\bibitem{jahangeer2011numerical}
K.~A. Jahangeer, A.~A.~O. Tay, and M.~R. Islam, ``Numerical investigation of
  transfer coefficients of an evaporatively-cooled condenser,'' \emph{Appl.
  Therm. Eng.}, vol.~31, no.~10, pp. 1655--1663, 2011.

\bibitem{ruz2017hybrid}
M.~L. Ruz, J.~Garrido, F.~V{\'a}zquez, and F.~Morilla, ``A hybrid modeling
  approach for steady-state optimal operation of vapor compression
  refrigeration cycles,'' \emph{Appl. Therm. Eng.}, vol. 120, pp. 74 -- 87,
  2017.

\bibitem{Bejarano2017}
G.~Bejarano, J.~A. Alfaya, M.~G. Ortega, and M.~Vargas, ``On the difficulty of
  globally optimally controlling refrigeration systems,'' \emph{Appl. Therm.
  Eng.}, vol. 111, pp. 1143--1157, 2017.

\bibitem{bejarano2017suboptimal}
G.~Bejarano, C.~Vivas, M.~G. Ortega, and M.~Vargas, ``Suboptimal hierarchical
  control strategy to improve energy efficiency of vapour-compression
  refrigeration systems,'' \emph{Appl. Therm. Eng.}, vol. 125, pp. 165--184,
  2017.

\bibitem{rubio2018optimal}
F.~R. Rubio, S.~J. Navas, P.~Ollero, J.~M. Lemos, and M.~G. Ortega, ``Optimal
  control applied to distributed solar collector fields,'' \emph{Rev. Iberoam.
  de Automat. e Inform. Ind.}, vol.~15, pp. 327--338, 2018.

\bibitem{navas2018optimal}
S.~J. Navas, F.~R. Rubio, P.~Ollero, and J.~M. Lemos, ``Optimal control applied
  to distributed solar collector fields with partial radiation,'' \emph{Sol.
  Energy}, vol. 159, pp. 811--819, 2018.

\bibitem{dincer2002bthermal}
I.~Dincer, ``On thermal energy storage systems and applications in buildings,''
  \emph{Energy and Build.}, vol.~34, no.~4, pp. 377--388, 2002.

\bibitem{maccracken2004thermal}
M.~M. MacCracken, ``Thermal energy storage myths,'' \emph{Energy Eng.}, vol.
  101, no.~4, pp. 69--80, 2004.

\bibitem{rismanchi2012energy}
B.~Rismanchi, R.~Saidur, G.~BoroumandJazi, and S.~Ahmed, ``Energy, exergy and
  environmental analysis of cold thermal energy storage {(CTES)} systems,''
  \emph{Renew. and Sustain. Energy Rev.}, vol.~16, no.~8, pp. 5741--5746, 2012.

\bibitem{mehling2008heat}
H.~Mehling and L.~F. Cabeza, \emph{Heat and cold storage with {PCM}}.\hskip 1em
  plus 0.5em minus 0.4em\relax Springer, 2008.

\bibitem{oro2012review}
E.~Or{\'o}, A.~De~Gracia, A.~Castell, M.~Farid, and L.~Cabeza, ``Review on
  phase change materials ({PCMs}) for cold thermal energy storage
  applications,'' \emph{Appl. Energy}, vol.~99, pp. 513--533, 2012.

\bibitem{verma2008review}
P.~Verma and S.~Singal, ``Review of mathematical modeling on latent heat
  thermal energy storage systems using phase-change material,'' \emph{Renew.
  and Sustain. Energy Rev.}, vol.~12, no.~4, pp. 999--1031, 2008.

\bibitem{dutil2011review}
Y.~Dutil, D.~R. Rousse, N.~B. Salah, S.~Lassue, and L.~Zalewski, ``A review on
  phase-change materials: mathematical modeling and simulations,'' \emph{Renew.
  and Sustain. Energy Rev.}, vol.~15, no.~1, pp. 112--130, 2011.

\bibitem{Bejarano2017NovelSchemePCM}
G.~Bejarano, J.~J. Suffo, M.~Vargas, and M.~G. Ortega, ``Novel scheme for a
  {PCM}-based cold energy storage system. {Design}, modelling and simulation,''
  \emph{Appl. Therm. Eng.}, vol. 132, pp. 256 -- 274, 2018.

\bibitem{bejarano2018efficient}
G.~Bejarano, M.~Vargas, M.~G. Ortega, F.~Casta{\~n}o, and J.~E. Normey-Rico,
  ``Efficient simulation strategy for {PCM}-based cold-energy storage
  systems,'' \emph{Appl. Therm. Eng.}, vol. 139, pp. 419 -- 431, 2018.

\bibitem{CoolProp}
\BIBentryALTinterwordspacing
I.~H. Bell, J.~Wronski, S.~Quoilin, and V.~Lemort, ``Pure and pseudo-pure fluid
  thermophysical property evaluation and the open-source thermophysical
  property library {CoolProp},'' \emph{Ind. and Eng. Chem. Res.}, vol.~53,
  no.~6, pp. 2498--2508, 2014. [Online]. Available: \url{www.coolprop.org}
\BIBentrySTDinterwordspacing

\bibitem{GB_JE_2015}
G.~Bejarano, J.~A. Alfaya, M.~G. Ortega, and F.~R. Rubio, ``Design, automation
  and control of a two-stage, two-load-demand experimental refrigeration
  plant,'' in \emph{$23^{rd}$ Mediterr. Conf. on Control and Autom.,
  Torremolinos (Spain)}, 2015, pp. 537--544.

\bibitem{bejarano2015multivariable}
G.~Bejarano, J.~A. Alfaya, M.~G. Ortega, and F.~R. Rubio, ``Multivariable
  analysis and {$H_\infty$} control of a one-stage refrigeration cycle,''
  \emph{Appl. Therm. Eng.}, vol.~91, pp. 1156--1167, 2015.

\bibitem{bejarano2016identifying}
G.~Bejarano, D.~Rodr{\'\i}guez, J.~A. Alfaya, M.~G. Ortega, and F.~Casta{\~n}o,
  ``On identifying steady-state parameters of an experimental
  mechanical-compression refrigeration plant,'' \emph{Appl. Therm. Eng.}, vol.
  109, pp. 318--333, 2016.

\bibitem{rodriguez2017parameter}
D.~Rodr{\'\i}guez, G.~Bejarano, J.~A. Alfaya, M.~G. Ortega, and F.~Casta{\~n}o,
  ``Parameter identification of a multi-stage, multi-load-demand experimental
  refrigeration plant,'' \emph{Control Eng. Pract.}, vol.~60, pp. 133--147,
  2017.

\bibitem{McKinley}
T.~L. McKinley and A.~G. Alleyne, ``An advanced nonlinear switched heat
  exchanger model for vapor compression cycles using the moving-boundary
  method,'' \emph{Int. J. of Refrig.}, vol.~31, no.~7, pp. 1253--1264, 2008.

\bibitem{BINLI}
B.~Li and A.~G. Alleyne, ``A dynamic model of a vapor compression cycle with
  shut-down and start-up operations,'' \emph{Int. J. of Refrig.}, vol.~33,
  no.~3, pp. 538--552, 2010.

\bibitem{pangborn2015comparison}
H.~Pangborn, A.~G. Alleyne, and N.~Wu, ``A comparison between finite volume and
  switched moving boundary approaches for dynamic vapor compression system
  modeling,'' \emph{Int. J. of Refrig.}, vol.~53, pp. 101--114, 2015.

\bibitem{Shen2010}
Y.~Shen, W.-J. Cai, and S.~Li, ``Normalized decoupling control for
  high-dimensional {MIMO} processes for application in room temperature control
  {HVAC} systems,'' \emph{Control Eng. Pract.}, vol.~18, no.~6, pp. 652--664,
  2010.

\bibitem{rodriguez2018robust}
D.~Rodr{\'\i}guez, G.~Bejarano, J.~A. Alfaya, and M.~G. Ortega, ``Robust and
  decoupling approach to {PID} control of vapour-compression refrigeration
  systems,'' in \emph{$3^{rd}$ IFAC Conf. on Adv. in
  Proportional-Integral-Derivative Control, Ghent (Belgium)}.\hskip 1em plus
  0.5em minus 0.4em\relax IFAC, 2018, pp. 698--703.

\bibitem{bejarano2018optimization}
G.~Bejarano, D.~Rodr{\'\i}guez, J.~A. Alfaya, J.~D. Gil, and M.~G. Ortega,
  ``Optimization and cascade robust temperature control of a refrigerated
  chamber,'' in \emph{$9^{th}$IFAC Symp. on Robust Control Des., Florianopolis
  (Brazil)}, 2018, pp. 110--115.

\bibitem{Bristol1966}
E.~H. Bristol, ``On a new measure of interaction for multivariable process
  control,'' \emph{IEEE Trans. on Autom. Control}, vol.~11, pp. 133--134, 1966.

\end{thebibliography}

\end{document}